%% file: ms.tex
\documentclass[preprint,3p]{elsarticle}

\usepackage{graphicx}
\usepackage{caption}
\usepackage{subcaption}
\usepackage{comment}
\usepackage{amssymb}
\usepackage{amsmath}
\usepackage{dirtytalk}
\usepackage{algorithm}
\usepackage{algorithmic}
\biboptions{numbers,sort&compress}
\usepackage{lineno}

\usepackage{xcolor}
\usepackage{hyperref}
\usepackage{siunitx}
\usepackage{tikz}
\usepackage{setspace}

\usepackage{pgfplots}
\usetikzlibrary{matrix}
\usetikzlibrary{plotmarks}
\pgfplotsset{compat=newest}
\usepackage{makecell}
\usepackage[export]{adjustbox}

\usepackage{amsmath}

\hypersetup{      
    colorlinks=true,                
    linkcolor= blue,                
    filecolor=black,
    citecolor=blue,
    linkbordercolor=blue
}
\graphicspath{ {./Figures/} }
\definecolor{brandeisblue}{rgb}{0.0, 0.44, 1.0}

\journal{Elsevier}
\begin{document}

\begin{frontmatter}


\title{An eXtended Finite Element Method Implementation in COMSOL Multiphysics: Solid Mechanics}




\author[civilUNSW]{Ahmad Jafari}
\author[civilShiraz]{Pooyan Broumand}
\author[civilUNSW]{Mohammad Vahab*}
\cortext[cor]{Corresponding author:}
\ead{m.vahab@unsw.edu.au}
\author[civilUNSW]{Nasser Khalili}

\address[civilUNSW]{School of Civil and Environmental Engineering, The University of New South Wales, Sydney 2052, Australia}
\address[civilShiraz]{Department of Civil and Environmental Engineering,  Shiraz University, Shiraz, Iran}

\begin{abstract}
This paper presents the first time implementation of the eXtended Finite Element Method (XFEM) in the general purpose commercial software COMSOL Multiphysics. An enrichment strategy is proposed, consistent with the structure of the software. To this end, for each set of enrichment functions, an additional Solid Mechanics module is incorporated into the numerical framework, coupled with compatible modifications to the internal variables. The Linear Elastic Fracture Mechanics (LEFM) is exclusively adopted for the crack analysis. The model pre-processing, level set update, stress intensity factor calculation and crack propagation analysis are conducted by employing COMSOL's built-in features in conjunction with external MATLAB functions through COMSOL LiveLink. All implementational aspects and suggested remedies for the treatment of enriched  elements,  framework  setup,  evaluation  of stress intensity factors, and numerical integration are described in detail. The accuracy and robustness of the proposed method are examined by several numerical examples for stationary and propagating crack problems in 2D and 3D settings. The results represent excellent agreement with available analytical, numerical and experimental observations in the literature.

\end{abstract}


\end{frontmatter}

\textbf{Highlights}

\begin{itemize}
\item XFEM implementation in COMSOL Multiphysics is proposed for the first time.
\item A straight-forward enrichment procedure is proposed with enormous potential for extension to multi-field problems.
\item Benchmark problems regarding single/multiple stationary/propagating cracks  in 2D/3D settings are explored and validated.
\end{itemize}

\input{introduction.tex}

\input{XFEMformulation.tex}
\input{COMSOLimplementation.tex}
\input{results.tex}
\input{conclusions.tex}


\bibliographystyle{model1-num-names}
\bibliography{references.bib}

\end{document}

%% file: introduction.tex
\section{Introduction}
\label{S:1 (introduction)}

Since its inception in 1999 by Belytschko and collaborators \cite{belytschko1999elastic,moes1999finite,daux2000arbitrary,dolbow2000discontinuous}, eXtended Finite Element Method (XFEM) has emerged as a versatile and rigorous computational tool for tackling \textit{weak/strong} discontinuities as well as high-gradients (i.e., \textit{singularities}). In XFEM, the special characteristics of the solution field are incorporated into the approximation space by means of the so-called enrichment functions. With the aid of the partition of unity concept, special enrichment functions are added to the standard approximation space associated with the classical finite element description \cite{khoei2014extended}. The mathematical foundation of the enrichment strategy to enhance the solution field traces back to the partition of unity finite element method (PUFEM) and the generalized finite element method (GFEM)  contributions (e.g., see Melenk and Babuška \cite{melenk1996partition}, Strouboulis et al. \cite{strouboulis2000design}), in which enrichment functions are employed at a universal level, contrary to XFEM where enrichments are utilized locally. Early contributions in XFEM were focused on the crack growth problem to demonstrate its performance in circumventing the need for remeshing, mesh refinements and data transfer (see Mohammadi \cite{mohammadi2008extended}). XFEM is now regarded as a proven technique in dealing with a broad range of applications, including linear elastic fracture mechanics (LEFM) (Moës et al. \cite{moes1999finite}, Sukumar et al. \cite{sukumar2000extended}, Chen et al. \cite{chen2012extended}), cohesive fractures (Zi and Belytschko \cite{zi2003new}, de'Borst et al.\cite{de2006mesh}), composite materials (Sukumar et al. \cite{sukumar2004partition}, Gracie and Belytschko \cite{gracie2009concurrently}, Akhondzadeh et al. \cite{akhondzadeh2017efficient}, Karimi et al. \cite{karimi2019adapting}, Pike and Oskay\cite{pike2015xfem}), shear band localization (Mikaeili and Schrefler \cite{mikaeili2018xfem}), contact mechanics (Liu et al. \cite{liu2008contact}, Broumand et al. \cite{BROUMAND201397}, Hirmand et al. \cite{hirmand2015augmented}), fluid–structure interaction (Legay et al. \cite{legay2006eulerian}), fractured porous media (de'Borst et al. \cite{de2006numerical}, Khoei et al. \cite{khoei2014mesh,khoei2018enriched}, Mohammadnejad and Khoei \cite{mohammadnejad2013extended}, Jafari et al. \cite{jafari2021fully}), and thermo-hydro-mechanical coupling processes (Khoei et al. \cite{khoei2012thermo}, Salimzadeh and Khalili \cite{salimzadeh2016fully}, Parchei and Gracie \cite{parchei2019undrained}), to name a few.

The increasing interests shown by both the computational mechanics communities and engineering end users have led into a variety of developments dedicated to open-access XFEM simulators. The notable examples include the open-source XFEM implementations by Sukumar and Prévost \cite{sukumar2003modeling}, who developed a Fortran implementation, and Dunant et al. \cite{dunant2007architecture}, who established an object-oriented programming library, for XFEM. Nonetheless, these and other similar in-house simulators often lack computational efficiency, a key ingredient in real-world engineering applications which commonly involve complex geometries, three dimensional settings, and extensive heterogeneities. As a remedy, there is a growing trend for the implementation of XFEM in general-purpose FE softwares featuring efficient built-in solvers and advanced meshing tools, such as ABAQUS, which permit new developments through addition of user defined subroutines. The substructuring approach to XFEM implementation in commercial packages, with no need to modification of the kernel, was first suggested by Wyart et al. \cite{wyart2008substructuring}. Giner et al. \cite{giner2009abaqus} employed user subroutine feature in ABAQUS (i.e., UEL) to simulate elastic fracture growth. Further improvements in relation to ABAQUS implementation of XFEM has been due to contributions by Cruz et al. \cite{cruz2019xfem} and Dehghan et al. \cite{dehghan20173d}, for intersecting fractures, by Xu and Yuan \cite{xu2009damage} and  Haddad and Sepehrnoori \cite{haddad2016xfem}, for cohesive fractures, and by Ooi et al. \cite{ooi2018investigating}, regarding contact mechanics.

In recent years, there has been an overwhelming demand for the elaboration of XFEM into multi-physics problems involving chemo-thermo-hydro-mechanical coupling analysis (e.g., see Khoei et al. \cite{khoei2012thermo}, Vahab et al. \cite{vahab2019x}, de'Borst et al. \cite{de2006numerical}). The inclusive capabilities of “COMSOL Multiphysics” in dealing with the simulation of multi-field problems and its attraction among researchers and engineers have been the incentive in this work to pursue the first XFEM implementation of COMSOL. A straightforward procedure is presented for the proposed implementation of XFEM by exploiting COMSOL's built-in features endowed with necessary external MATLAB functions, which can be clustered in the following tasks:\\
i) Adopt a compatible XFEM formulation according to the structure of COMSOL;\\
ii) Regenerate and modify the generic Solid Mechanics module of COMSOL to account for the presence of crack interfaces;\\
iii) Conduct the level set analysis, for tracking the interfaces, via external MATLAB functions to overcome the software's restriction in accessing data at nodal/elemental level prior to (i.e., at pre-processing stage) and during the analysis; and,\\
iv) Perform SIF evaluation by taking advantage of the internal functions and variables at the post-processing stage.\\
 The procedure proposed is robust and enables handling of complex scenarios in cracked media in 2D/3D domains. While it is formulated for solid mechanics simulations, it is eminently amenable to extension to multi-physics problems. 

The paper is organized as follows: In section \ref{S:2 (Formulation)}, the governing equations for XFEM formulation of fracture growth in an elastic domain are briefly described in conjunction with the weak forms and fracture growth criteria. Section \ref{S:6 (COMSOL impl)} is dedicated to the implementation of XFEM in COMSOL, which involves detailed algorithms employed for identification of the enriched elements, module setup, evaluation of the stress intensity factor, and numerical integration. In section \ref{S:4 (results)}, the performance of the proposed framework is investigated using a selection of benchmark examples, in 2D and 3D settings. Concluding remarks are presented in section \ref{S:5 (Conclusions)}. Transfer of the knowledge to academia and industry is a cornerstone of this paper, hence the proposed model is made available at \href{https://github.com/ahmadjafari93/xfem-comsol.git}{https://github.com/ahmadjafari93/xfem-comsol.git}.

%% file: XFEMformulation.tex
\section{XFEM formulation}
\label{S:2 (Formulation)}

In essence, XFEM decouples the interfaces, such as cracks or material discontinuities, from the background mesh by enriching the finite element space with special enrichment functions, based on the partition of unity method \cite{PUM1997}. Therefore, it eliminates the remeshing step which is required in the classical finite element modeling of moving interfaces. In this method, the handling of crack interface topology and its evolution are performed by using nodal distances to the corresponding projection points on the interface \cite{moes1999finite}. Alternatively, the Level Set Method (LSM) can be employed, for which the extension to higher dimensions and coupling with the XFEM is straightforward \cite{khoei2014extended}. The special treatment of the Galerkin finite element formulation, that is elaborated in the following section, facilitates the separation of the weak forms of the standard and enriched parts of the governing equations. The proposed formulation is inspired by the work of Borja et al. \cite{borja2008assumed} which is amenable to the modeling structure of COMSOL Multiphysics.

\subsection{Governing equations and XFEM discretization}
\label{S:2-1 govenrning&discretise}

\begin{figure}[!t]
\centering\includegraphics[width=0.65\linewidth]{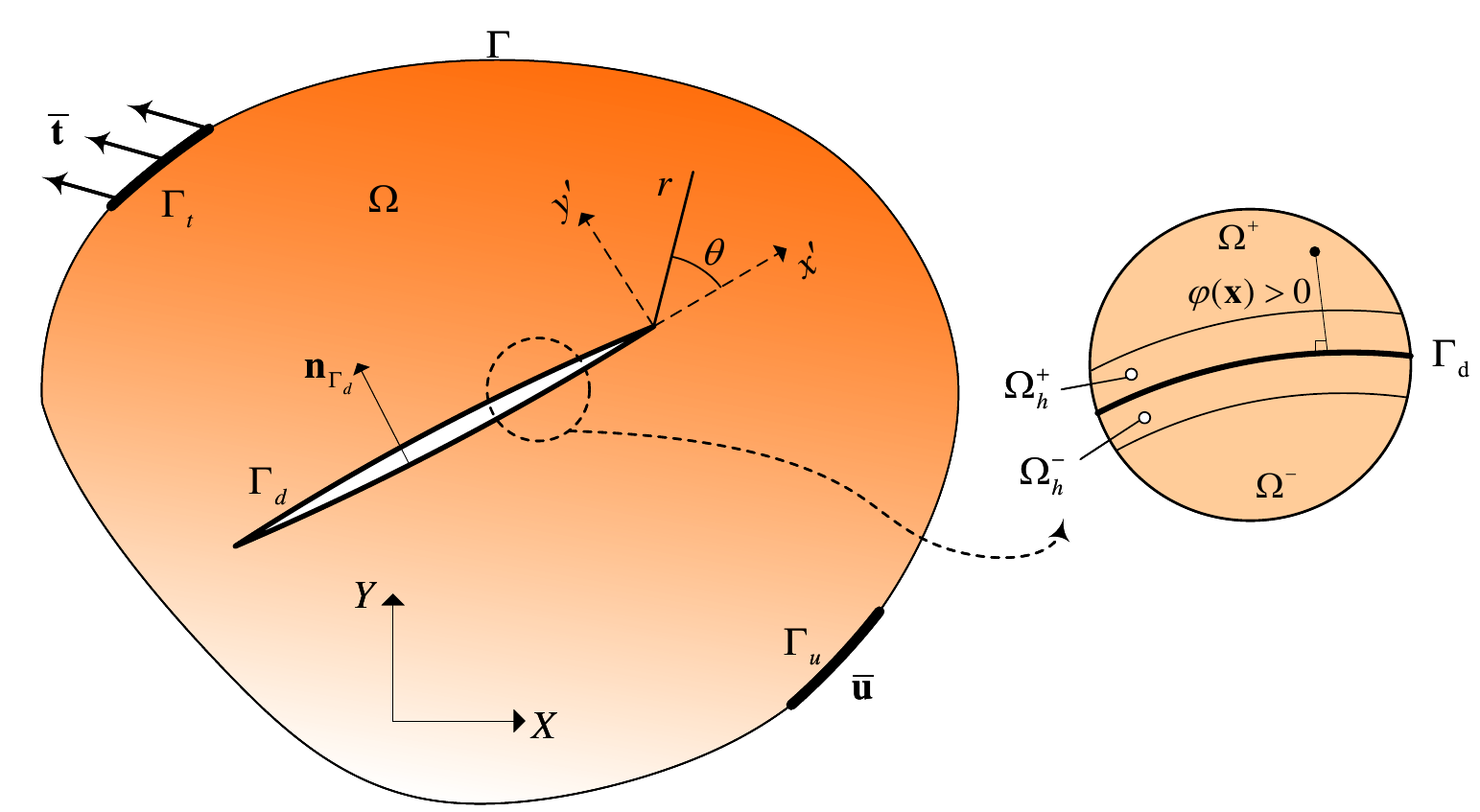}
\caption{Schematics of problem domain and boundaries of fractured media.}
\label{fig:potato}
\end{figure}

As shown in Fig. \ref{fig:potato}, consider a cracked body $\Omega$ that is bounded by $\Gamma =\Gamma _{u}\cup \Gamma _{t}$ and crack surfaces $\Gamma_d$, with $\Gamma _{u}\cap  \Gamma _{t}=\varnothing $. The equation of motion of the domain can be expressed as, 

\begin{equation}
\label{eq:equilibrium}
\begin{matrix}
\nabla \cdot \mathbf{\sigma} - \rho \ddot{\mathbf{u}} +\rho \mathbf{b}=0  & \text{in }\Omega
\end{matrix}
\end{equation}
subjected to the following boundary and initial conditions,
\begin{equation}
\label{eq:equilibrium BC}
\begin{matrix}
\mathbf{u}=\mathbf{\bar{u}} & \text{on }\Gamma _{u} \\ 
\mathbf{\sigma}\cdot \mathbf{n}_{\Gamma }=\mathbf{\bar{t}}& \text{on }\Gamma _{t} \\ 
\mathbf{\sigma}\cdot \mathbf{n}_{\Gamma_{\text{d}} }=\mathbf{\bar{t}}_{\text{d}}& \text{on }\Gamma _{\text{d}}\\
\mathbf{u}=\mathbf{u}_0 & \text{in }\Omega \\ 
\dot{\mathbf{u}}=\dot{\mathbf{u}}_0 & \text{in }\Omega \\ 
\end{matrix}
\end{equation}
where, $\rho$ is density and, $\mathbf{b}$, $\mathbf{u}$, $\dot{\mathbf{u}}$ and $\ddot{\mathbf{u}}$ are the body force, displacement, velocity and acceleration vectors, respectively. In this equation, $\mathbf{\bar{u}}$ and $\mathbf{\bar{t}}$ denote the prescribed displacement and traction vectors on the boundary of the domain and, $\mathbf{u}_0$ and $\dot{\mathbf{u}}_0$ are the initial displacement and  velocity vectors of the domain, respectively. $\mathbf{n}_{\Gamma }$ and $\mathbf{n}_{\Gamma _{\text{d}}}$ are the unit normal vectors to the external boundary and crack surfaces. $\mathbf{\sigma}$ is the Cauchy's stress tensor which is related to the strain tensor $\mathbf{\varepsilon}$  through Hooke’s law for isotropic elastic materials as $\mathbf{\sigma} = \mathbf{D}:\mathbf{\varepsilon}$, where $\mathbf{D}$ is the fourth order elasticity tensor.

As shown in Fig. \ref{fig:potato}, the displacement field is discontinuous across $\Gamma _{d}$, while the stress field is singular at the crack-tips; hence, in XFEM formulation it can be expressed as,

\begin{equation}
\label{eq:stdenr}
\mathbf{u}=\mathbf{u}^{\text{cont}}+ M_{\Gamma _{\text{d}}}(\mathbf{x}) \mathbf{u}^{\text{disc}}+\sum_{i=1}^{4}F_{\rm{i}}(\mathbf{x})\mathbf{u}_{\rm{i}}^{\text{tip}}
\end{equation}
where $\mathbf{u}^{\text{cont}}$, $M_{\Gamma _{\text{d}}}(\mathbf{x}) \mathbf{u}^{\text{disc}}$ and $\sum_{i=1}^{4}F_{\rm{i}}(\mathbf{x})\mathbf{u}_{\rm{i}}^{\text{tip}}$ are the continuous, discontinuous and crack tip terms associated with the displacement field. $M_{\Gamma _{\text{d}}}(\mathbf{x})$ is the shifted Heaviside enrichment function that generates discontinuity across $\Gamma _{d}$ by $M_{\Gamma _{\text{d}}}(\mathbf{x}) = H_{\Gamma _{d}}(\varphi (\mathbf{x}))=\mathbb{H}_{\Gamma _{d}}(\varphi (\mathbf{x}))-\mathbb{H}_{\Gamma _{d}}(\varphi (\mathbf{x^{I}}))$ \cite{liu2008contact}, where

\begin{equation}
\label{eq:Heavisied}
\mathbb{H}_{\Gamma _{d}}(\varphi (\mathbf{x}))=\left\{\begin{matrix}
1 & \varphi (\mathbf{x})\geq 0\\ 
-1 & \varphi (\mathbf{x})< 0
\end{matrix}\right.
\end{equation}
In the above relation, $\varphi (\mathbf{x})$ is the signed distance function corresponding to the discontinuity $\Gamma _{d}$, which is used to determine the enriched nodes and associated elements (see Fig. \ref{fig:potato}). Also, $F(\mathbf{x})=\left \{ F_{1},F_{2},F_{3},F_{4} \right \}$ is the set of asymptotic crack tip enrichment functions which are adopted from the analytical solutions of the crack tip process zone.

Considering Eq. \ref{eq:stdenr} for the discrete form of the displacement field, the infinitesimal strain tensor can be expressed as

\begin{equation}
\label{eq:epsdefine}
\mathbf{\varepsilon }=\nabla^{\text{s}}\mathbf{u}=\nabla^{\text{s}}\mathbf{u}^{\text{cont}}+ H_{\Gamma _{d}}(\varphi (\mathbf{x}))\nabla^{\text{s}}\mathbf{u}^{\text{disc}}+ \delta _{\Gamma _{\text{d}}} (\mathbf{u}^{\text{disc}}\otimes \mathbf{n_{\Gamma _{\text{d}}}} )^{\text{s}}+\nabla^{\text{s}}(\sum_{i=1}^{4}F_{\rm{i}}(\mathbf{x})\mathbf{u}_{\rm{i}}^{\text{tip}})
\end{equation}
where $\nabla^{\text{s}}$ and $(\cdot )^{\text{s}}$ denote the symmetric parts of the spatial gradient operator and tensor, respectively, and $\delta _{\Gamma _{\text{d}}}$ is the Dirac's delta function on $\Gamma_d$. In order to derive the weak form of Eq. \ref{eq:equilibrium}, a costume tailored test function $\mathbf{\eta }$ which is consistent with the displacement field is adopted as $\mathbf{\eta}=\mathbf{\eta}^{\text{cont}}+ M_{\Gamma _{\text{d}}}(\mathbf{x}) \mathbf{\eta}^{\text{disc}}+\sum_{i=1}^{4}F_{\rm{i}}(\mathbf{x})\mathbf{\eta}_{\rm{i}}^{\text{tip}}$. Following the standard approach in the calculus of variations, the weak form of the equation of motion is obtained as

\begin{equation}
\label{eq:vartotal}
\int_{\Omega }^{}\nabla^{\text{s}}\mathbf{\eta}:\mathbf{\sigma}\text{d}\Omega = \int_{\Omega }^{}\mathbf{\eta}\cdot \rho \mathbf{b}\text{d}\Omega - \int_{\Omega }^{}\mathbf{\eta}\cdot \rho \ddot{\mathbf{u}}\text{d}\Omega +\int_{\Gamma _{\text{t}}}^{}\mathbf{\eta}\cdot \mathbf{\bar{t}}\text{d}\Gamma 
\end{equation}
Substituting $\mathbf{\eta }$ in form of two independent weight functions $\mathbf{\eta}^{\text{cont}}$ and $\mathbf{\eta}^{\text{disc}}$  into Eq. \ref{eq:vartotal}, the weak form of the continuous part of the governing equations yields as

\begin{equation}
\label{eq:varstd}
\int_{\Omega }^{}\nabla^{\text{s}}\mathbf{\eta}^{\text{cont}}:\mathbf{\sigma}\text{d}\Omega = \int_{\Omega }^{}\mathbf{\eta}^{\text{cont}} \cdot \rho \mathbf{b}\text{d}\Omega - \int_{\Omega }^{}\mathbf{\eta}^{\text{cont}}\cdot \rho \ddot{\mathbf{u}}\text{d}\Omega +\int_{\Gamma _{\text{t}}}^{}\mathbf{\eta}^{\text{cont}} \cdot \mathbf{\bar{t}}\text{d}\Gamma 
\end{equation}
and, the discontinuous and singular tip-enrichment parts can be expressed as
\begin{equation}
\label{eq:varenr}
\begin{split}
\int_{\Omega }^{}[H_{\Gamma _{d}}(\varphi (\mathbf{x}))\nabla^{\text{s}}\mathbf{\eta}^{\text{disc}}]:\mathbf{\sigma}\text{d}\Omega +  \int_{\Gamma _{\text{d} }}^{}\mathbf{\eta}^{\text{disc}} \mathbf{\sigma}\cdot \mathbf{n}_{\Gamma _{\text{d}}}\text{d}\Gamma= &\int_{\Omega }^{}H_{\Gamma _{d}}(\varphi (\mathbf{x}))\mathbf{\eta}^{\text{disc}} \cdot \rho \mathbf{b}\text{d}\Omega\\
& - \int_{\Omega }^{}H_{\Gamma _{d}}(\varphi (\mathbf{x}))\mathbf{\eta}^{\text{disc}} \cdot \rho \ddot{\mathbf{u}}\text{d}\Omega
 +\int_{\Gamma _{\text{t}}}^{}H_{\Gamma _{d}}(\varphi (\mathbf{x}))\mathbf{\eta}^{\text{disc}} \cdot \mathbf{\bar{t}}\text{d}\Gamma
\end{split}
\end{equation}

\begin{equation}
\label{eq:vartip}
\int_{\Omega }^{}\nabla^{\text{s}}(\sum_{i=1}^{4}F_{\rm{i}}(\mathbf{x})\mathbf{\eta}_{\rm{i}}^{\text{tip}}):\mathbf{\sigma}\text{d}\Omega = \int_{\Omega }^{}\sum_{i=1}^{4}F_{\rm{i}}(\mathbf{x})\mathbf{\eta}_{\rm{i}}^{\text{tip}} \cdot \rho \mathbf{b}\text{d}\Omega - \int_{\Omega }^{}\sum_{i=1}^{4}F_{\rm{i}}(\mathbf{x})\mathbf{\eta}_{\rm{i}}^{\text{tip}}\cdot \rho \ddot{\mathbf{u}}\text{d}\Omega 
\end{equation}
The integration domain of Eq. \ref{eq:varenr} and Eq. \ref{eq:vartip} are limited to the supports of $M_{\Gamma _{\text{d}}}(\mathbf{x})$ and $F(\mathbf{x})$, which are the enriched zone detected by the signed distance function $\Omega _{\text{h}}$ and asymptotic crack tip function $\Omega _{\rm{tip}}$, respectively. Since in this study, cracks are stipulated as traction free, the second term in Eq. \ref{eq:varenr} vanishes.

Adopting a Galerkin formulation, the trial and test functions are discretized by $C^{0}$ continuous shape functions $N_{\text{\rm{i}}} (\mathbf{x})$ which are associated with the vector of nodal displacements for standard $\hat{\mathbf{u}}$ and enriched parts including discontinuous $\tilde{\mathbf{u}}$ and crack tip $\bar{\mathbf{u}}$ as

\begin{equation}
\label{eq:descretise}
\left\{ {\begin{array}{*{20}{l}}
{{{\bf{u}}^{{\rm{cont}}}}({\bf{x}}) = \sum\nolimits_{{\rm{i}} \in {m_{{\rm{std}}}}} {{N_{\rm{i}}}({\bf{x}}){\bf{\hat u}_{\rm{i}}}} }&{{\rm{in}}\,\Omega }\\
{{{\bf{u}}^{{\rm{disc}}}}({\bf{x}}) = \sum\nolimits_{{\rm{i}} \in {m_{{\rm{disc}}}}} {{N_{\rm{i}}}({\bf{x}}){H_{{\Gamma _{\rm{d}}}}}(\varphi ({\bf{x}})){\bf{\tilde u}_{\rm{i}}}} }&{{\rm{in}}\,{\Omega _{\rm{h}}}}\\
{{{\bf{u}}^{{\rm{tip}}}}({\bf{x}}) = \sum\nolimits_{{\rm{i}} \in {m_{{\rm{tip}}}}} {{N_{\rm{i}}}({\bf{x}})\sum_{\rm{i}=1}^{4}F_{\rm{j}}(\mathbf{x}){\bf{\bar u}_{\rm{ij}}}} }&{{\rm{in}}\,{\Omega _{\rm{tip}}}}
\end{array}} \right.
\end{equation}
where $m_\text{std}$, $m_\text{disc}$ and $m_\text{tip}$ are sets of standard, discontinuous and tip enrichment nodes, respectively.

\subsection{Fracture criteria and crack propagation}
\label{S:3-2 SIF}

The interaction integral method is an effective energy approach which is based on $J$-integral concept, and it is widely used in the calculation of mixed-mode stress intensity factors (SIFs) \cite{anderson2017fracture}. This method takes advantage of auxiliary fields, available from analytical solutions, that is superimposed on the calculated fields. Typically, boundary or domain form of the interaction integral is used to evaluate the stress intensity factors as a post-process. The energy release rate of a solid body in two dimensions is expressed as 

\begin{equation}
\label{eq:J1}
J=\frac{K_{I}^{2}+K_{II}^{2}}{{E}'}
\end{equation}
where ${E}'$ is defined as $E/(1-\upsilon ^2)$ and $E$ for plane strain and plane stress problems, respectively. The contour form of the $J$-integral is represented as
\begin{equation}
\label{eq:J2}
J=\int_{\Gamma _{J}}^{}\left [ w\cdot n_{{x}'}-(\mathbf{\sigma \cdot \nabla_{{x}'}\mathbf{u}})\cdot \mathbf{n} \right ]d\Gamma 
\end{equation}
where $w$ is the strain energy density function, $\mathbf{n}$ and $n_{{x}'}$ are the unit normal vector and its horizontal component (with respect to local crack coordinates ${x}'-{y}'$) of the closed curved path $\Gamma _{J}$ encompassing the crack-tip, respectively. $\nabla_{{x}'}$ is the directional gradient operator in the local horizontal direction. Applying Eq. \ref{eq:J2} to the actual and auxiliary fields, the interaction integral takes the form

\begin{equation}
\label{eq:I12}
I^{(1+2)}=\int_{\Gamma _{J}}^{}\left [ W^{(1,2)}\cdot n_{{x}'}-(\mathbf{\sigma }^{{(1)}}\nabla_{{x}'}\mathbf{u}^{(2)}+\mathbf{\sigma }^{{(2)}}\nabla_{{x}'}\mathbf{u}^{(1)})  \right ]d\Gamma 
\end{equation}
in which superscripts $(1)$ and $(2)$ respectively represent the actual and auxiliary states, and $W^{(1,2)} = \mathbf{\sigma }^{(1)}\cdot \mathbf{\varepsilon }^{(2)}=\mathbf{\sigma }^{(2)}\cdot \mathbf{\varepsilon }^{(1)}$ is the interaction strain energy. Combining Eqs.  \ref{eq:J1}, \ref{eq:J2} and \ref{eq:I12}, it can be concluded that

\begin{equation}
\label{eq:I12K}
I^{(1+2)}=\frac{2}{{E}'}(K_{\text{I}}^{(1)}K_{\text{I}}^{(2)}+ K_{\text{II}}^{(1)}K_{\text{II}}^{(2)})
\end{equation}
By appropriate selection of auxiliary fields for pure mode I (i.e., $K_{I}^{(2)}=1$, $K_{II}^{(2)}=0$) and mode II (i.e., $K_{I}^{(2)}=0$, $K_{II}^{(2)}=1$), the stress intensity factors of mixed-mode problems can be calculated. In addition, a domain form of Eq. \ref{eq:I12} can be obtained by application of Gauss-divergence theorem and use of special weighting functions (see Anderson \cite{anderson2017fracture}).

In order to estimate the fracture propagation direction, the maximum hoop stress criteria \cite{mohammadi2008extended,giner2009abaqus} is employed. Based on the calculated values of SIFs, the propagation angle $\theta _{c}$ is obtained as
\begin{equation}
\label{eq:thetac}
\theta _{c}= \text{cos}^{-1}\left ( \frac{3K_{\text{II}}^{2} + \sqrt{K_{\text{I}}^{4}+8K_{\text{I}}^{2}K_{\text{II}}^{2}}}{K_{\text{I}}^{2}+9K_{\text{II}}^{2}} \right )
\end{equation}
where $\theta _{c}$ is measured with respect to the current local coordinate system of the associated crack tip. Using the propagation angle $\theta _{c}$, an arbitrary crack increment is added to the existing crack configuration, and the solution continues. For more information on XFEM implementation of fractures and issues related to blending elements, refer to \cite{khoei2014extended, mohammadi2008extended}.

%% file: COMSOLimplementation.tex
\section{COMSOL implementation}
\label{S:6 (COMSOL impl)}

\subsection{Overview}
\label{S:3-1 overview}

COMSOL Multiphysics is a general-purpose simulation software for multi-field problems, that is based on the finite element method. In this software, multiple physics can be combined by employing the available built-in interfaces or implementing user defined physics \cite{multiphysics2019introduction}. The Solid Mechanics module offers the general formulation for the analysis of solid bulk based on the principles of continuum mechanics \cite{multiphysics2019structural}. Thus, in this approach, the  discontinuities in the solution domain must be explicitly modeled and the generated mesh must conform to the interface boundaries. To overcome this restriction, the proposed XFEM implementation in this study takes advantage of six distinct Solid Mechanics modules as per weak form equations developed in section \ref{S:2-1 govenrning&discretise}; i.e., one for the standard part of the displacement field $\mathbf{u}^{\text{cont}}$, one for the discontinuous enriched component $\mathbf{u}^{\text{disc}}$, and four other modules to account for the asymptotic tip enrichment part $\mathbf{u}^{\text{tip}}$. This is plausible by exploiting the interesting feature of COMSOL in provision of the access to the definitions of the field variables (e.g., stress and strain fields). For the sake of brevity, the Solid Mechanics module that deals with the standard displacement field is referred to as SMstd, the discontinuous enriched module is denoted by SMenr, and the crack tip enriched modules are indicated by SMtip in the rest of the work. This capability of COMSOL can be elaborated identically for XFEM developments in a wide range of problems with several enrichment functions associated with complex coupled physics such as thermo-hydro-mechanical coupling analysis. The XFEM implementation of multiple physics in COMSOL could be the subject of future studies.

COMSOL Multiphysics features several internal variables and functions (e.g., path/domain integration tools) which are critical to the successful execution of the proposed XFEM analysis. These options are primarily utilized to realize the enrichment concept as well as to perform the pre-processing (e.g., level set calculation) and post-processing (e.g., SIF calculation and crack propagation) functions. In this respect, the Live-link for MATLAB feature offers excellent flexibility to the developer to implement the required subroutines ground up. \cite{multiphysics2018matlab}.  

\subsection{Pre-processing of the crack geometry}
\label{S:3-2 Pre-process}

The pre-processing is performed for the identification of the crack geometry prior to the XFEM analysis. To this end, a conventional finite element mesh is generated in COMSOL. The mesh is exported as a text file using \say{*.mphtxt} format, in which the nodal coordinates and element connectivities are reported. The file is then imported into MATLAB as a script file, that is named \say{preprocess.m}, where the level set function is defined. In this code, a geometric search is carried out on all elements' nodes and edges to determine their 
position with respect to the existing crack interfaces. The output of the pre-processing phase is the list of all nodes and elements which their support domain is bisected by the crack interface, or contain the crack tips.  This is utilized to initialize the XFEM analysis as well as to determine the enriched zone $\Omega_{h}$, over which SMenr and SMtips modules are activated.

\begin{figure}[!t]
\centering\includegraphics[width=0.55\linewidth]{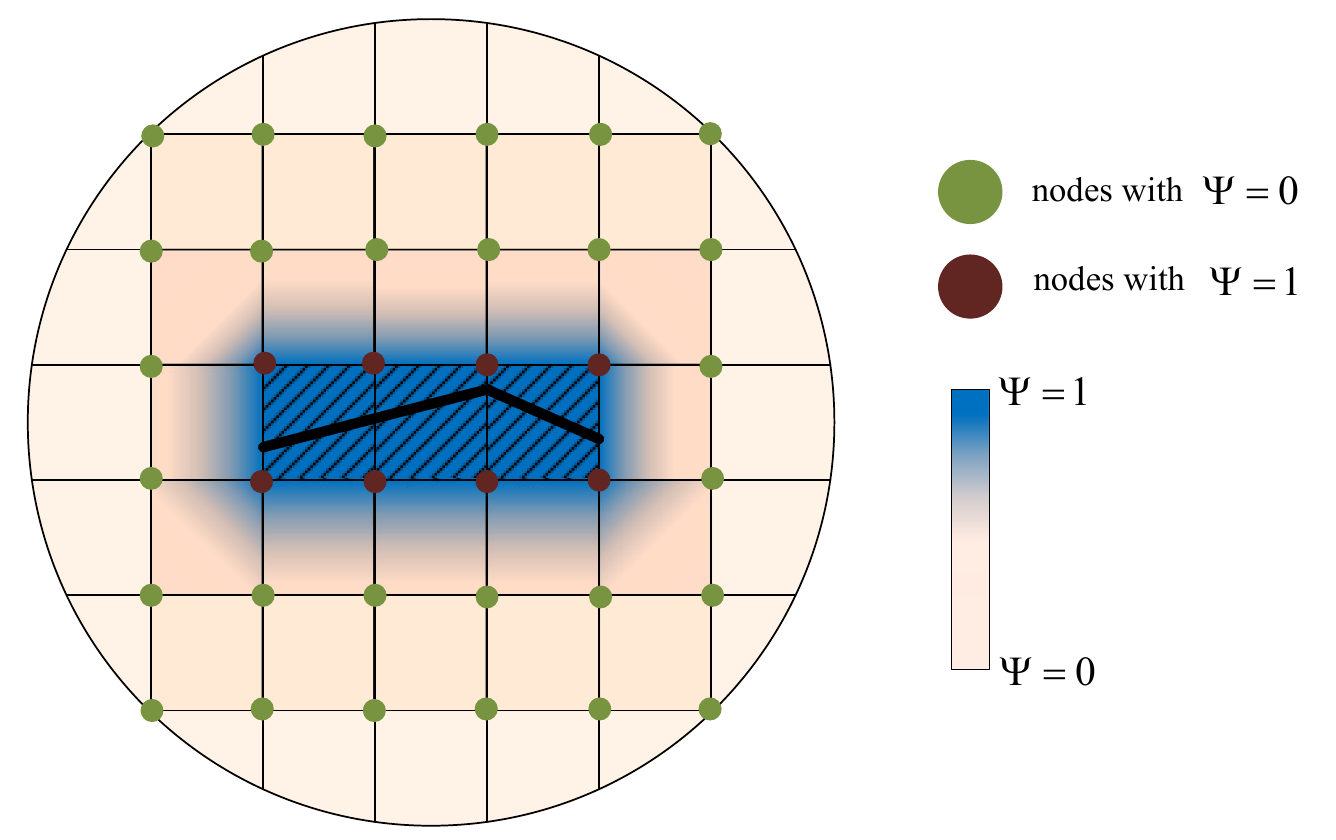}
\caption{Detection of fractured elements in COMSOL using interpolation of $\psi$ field variable.}
\label{fig:psi}
\end{figure}

The pre-processing procedure is repeated as per geometrical update of the crack interface. This is achieved via two MATLAB functions named \say{phi.m} and \say{interpol.m}. The former calculates the Heaviside function values for any arbitrary point of interest (e.g., a Gauss-point), which are used later to modify the strain field in SMenr module. The \say{interpol.m} function is utilized to detect the enriched zone by employing an interpolation function. Since COMSOL does not provide direct access to the elemental and nodal data of the model at any stage of the analysis, the interpolation function is required to infer such data geometrically from the original mesh . This function is set to be zero across the domain except at enriched nodes, for which it is equal to unity. For any point of interest, the output variable, that is called $\psi$, is interpolated using the MATLAB built-in function \textit{scatteredInterpolant} inside the \say{interpol.m} function. Fig. \ref{fig:psi} schematically represents the definition of $\psi$ over the domain. At the end of the pre-processing phase all nodal signs, element connectivities, crack tip coordinates and interpolation values of nodal points are saved in separate MAT-files. This facilitates the access to the data at anytime during the course of the analysis, for which a MATLAB function is called. It is worth noting that the number of defined MATLAB functions must match the number of variables required throughout the analysis. Furthermore, for each function being called, all input and output vectors must have the identical sizes \cite{multiphysics2018matlab}.

\subsection{Module setup: enrichment}
\label{S:3-2 module}

The six Solid Mechanics modules that represent the standard and enriched fields (i.e., SMstd, SMern and SMtips) are established based on the presented weak formulation in section \ref{S:2-1 govenrning&discretise}. In SMstd, the continuous part of the strain field is predefined as $\mathbf{\varepsilon }^{\text{cont}} = \nabla^{\text{s}}\mathbf{u }^{\text{cont}}$, with no modification  required. In contrast, to implement the strong discontinuities in SMenr module, the default definitions of displacement gradient and the associated strain field is modified by incorporating the enrichment function $H_{\Gamma _{d}}(\varphi (\mathbf{x}))$, which is obtained from the MATLAB function \say{phi.m}, as

\begin{equation}
\label{eq:epsenrimplement}
\mathbf{\varepsilon }^{\text{disc}}=\nabla^{\text{s}}\mathbf{u }^{\text{disc}}\times H_{\Gamma _{d}}(\varphi (\mathbf{x}))
\end{equation}
where $\mathbf{\varepsilon }^{\text{disc}}$ represents the discontinuous enriched part of the strain tensor. This modification is performed by enabling the \say{Equation View} option in \say{Model Builder} panel of COMSOL, that provides access to the definition of variables associated with the corresponding module. Similarly, the strain contribution from the asymptotic tip enrichments can be expressed as

\begin{equation}
\label{eq:epstipimplement}
\mathbf{\varepsilon }^{\text{tip}}=\nabla^{\text{s}}(\sum_{i=1}^{4}F_{\rm{i}}(\mathbf{x})\mathbf{u}_{\rm{i}}^{\text{tip}})
\end{equation}

In addition to the strain field, the definition of the stress field in all Solid Mechanics modules must be modified so as to ensure a unique stress field is reproduced over the whole solution domain. Consequently, the stress field $\mathbf{\sigma }^{\text{total}}$ (i.e., second Piola-Kirchhoff stress in COMSOL) in SMstd is modified as

\begin{equation}
\label{eq:sigmacomsol}
\mathbf{\sigma }^{\text{total}}=\mathbf{D}:({\mathbf{\varepsilon }^{\text{cont}}}+{\mathbf{\varepsilon }^{\text{disc}}}+{\mathbf{\varepsilon }^{\text{tip}}})
\end{equation}
Note that the stress field in the SMenr and SMtips are also identically set according to the relation in (\ref{eq:sigmacomsol}).

As described in section \ref{S:3-2 Pre-process}, SMenr and SMtips must be defined exclusively on the enriched zones of the domain i.e., $\Omega_h$. However, COMSOL restricts access to the nodal data. To circumvent this difficulty, in the approach presented herein, $\Omega_h$ is defined as a subset of $\Omega$ over which the predefined enriched degrees of freedom are not restrained; the modules for SMenr and SMtip are initialized based on the same geometry and background mesh as that of SMstd. This is effected by selecting \say{Prescribed Displacement} from \say{Domain Constraint} option of the module, that enables the imposition of a predefined displacement field to the domain. For this purpose, the field variable $\psi$, introduced in section \ref{S:3-2 Pre-process} (i.e., the output of \say{interpol.m} MATLAB function), is utilized to prescribe the displacements as 
\begin{equation}
\label{eq:interpolconstraint}
{{\bf{u}}^{{\rm{disc}}}} = \left\{ {\begin{array}{*{20}{l}}
{{{\bf{u}}^{{\rm{disc}}}}}&{{\rm{ where\: } } \psi {\rm{ = 1}}}\\
0&{{\rm{otherwise}}}
\end{array}} \right.
\end{equation}
The constraint in (\ref{eq:interpolconstraint}) eliminates the extra DOFs that are outside of the enriched zone in SMenr module. In this way, not only the computational costs are reduced significantly, but also the stiffness matrix singularity due to the presence of zero values in SMenr and SMtips is avoided; i.e. due to the \say{zero} extension of the enrichment function over the un-enriched zones of the domain.

\subsection{Stress intensity factors}
\label{S:3-3 sifs}
In this work, the stress intensity factors are calculated by employing COMSOL's internal variables in conjunction with the built-in mathematical operators and functions. The list of required internal variables is defined in a COMSOL script called \say{interaction integral} in the \say{Definitions} section, that includes the crack-tip coordinates, normal vector components, displacement, stress and strain derivatives, interaction strain energy density function, interaction integral and SIF values. 

The equivalent domain form of the interaction integral is typically used in the calculation of the SIFs in the literature \cite{anderson2017fracture}; however, the path integral form (i.e., Eq. \ref{eq:I12}) is preferred here for the sake of simplicity of the  application, and the availability of the built-in circular path integral operator in COMSOL, called \textit{circint}. In this respect, \textit{at2} operator is also used to set the center of the circular path to the crack tips, around which the interaction integral is calculated. Alternatively, the \textit{diskint} operator could be used for the calculation of domain form of the interaction integral along a circular area surrounding the crack-tip. 

\subsection{Numerical integration}
\label{S:3-4 (integration)}

In the classical FEM, piecewise continuous polynomials are used to discretize the displacement field, which are integrated accurately by relatively lower-order Gauss integration rules.  However, in XFEM, due to the existence of singularities and/or discontinuities in the displacement field and its derivatives, a more precise integration strategy is required for the enriched part of the displacement field. In this respect, raising the order of integration by increasing the number of Gauss-points, triangular/rectangular partitioning of the elements, and the rectangular sub-griding  are among the most frequently used methods in the literature \cite{mohammadi2008extended, khoei2014extended}. The last two methods are not available in COMSOL and therefore, the first approach is adopted in this study for SMenr and SMtips modules.

\begin{figure}[!t]
\centering\includegraphics[width=0.7\linewidth]{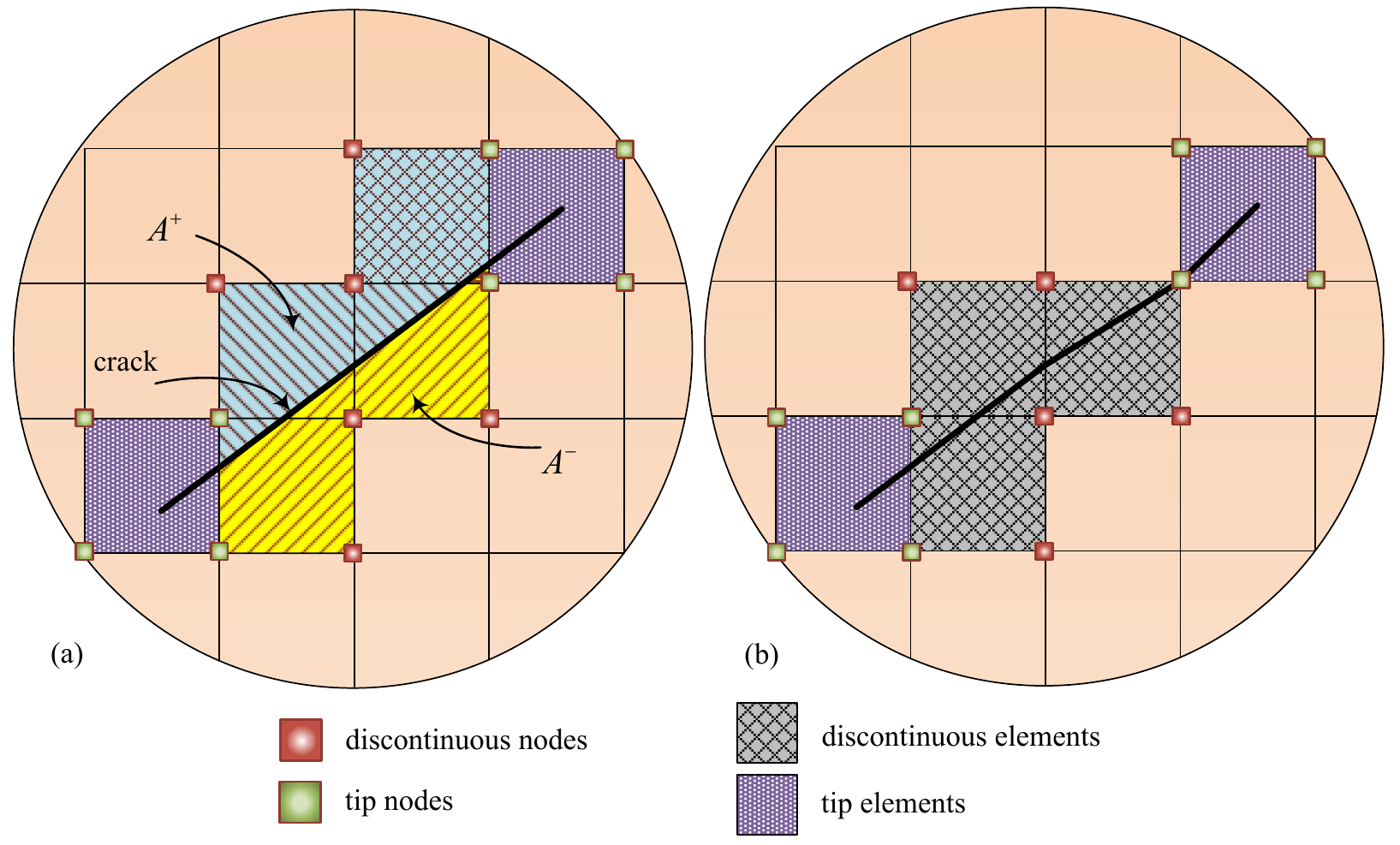}
\caption{ \textbf{a)} Definition of the support domain of a node for the enrichment criterion, \textbf{b)} enrichment modification: the criterion is met and some of the nodes are not enriched; this can happen when a crack is very close to a node point.}
\label{fig:effective area}
\end{figure}

\textbf{Remark 1.} In order to avoid ill-conditioned and/or singular stiffness matrices in XFEM, it is necessary to ensure that there exists at least a minimum number of Gauss points at either sides of the interface in the cracked elements. Hence, it is required to apply a criterion for the size of the support domain of each nodal point corresponding to a particular integration order (Fig. \ref{fig:effective area}a). In this respect, the nodes for which the relative support domain, i.e., the ratio $A^{+}/(A^{+}+A^{-})$ or $A^{-}/(A^{+}+A^{-})$, is smaller than a predefined tolerance $\delta$ are not enriched \cite{mohammadi2008extended} (see Fig. \ref{fig:effective area}b).

\subsection{Analysis and sequencing of processes}
\label{S:3-5 (sequencing)}

The numerical solution to the discretized form of the governing equations, represented in the preceding sections, is obtained by using the \say{Study} node in COMSOL. There are several study options in the software such as \say{Stationary} and \say{Time Dependent}, which correspond to quasi-static and dynamic analysis strategies, respectively. In the quasi-static analysis of XFEM problems, the prescribed load (or displacement) is applied incrementally; at each step of loading, the crack propagation criterion is checked, and a predefined increment is added to the crack interface, if the propagation criterion is met. In LEFM, the problem is linear both geometrically and from the material behaviour point of view. In order to avoid the issues related to data transfer following each crack propagation step, typically the problem is solved from the beginning yet with an updated configuration for the crack geometry (e.g., see \cite{khoei2014extended,BROUMAND201397}). This process can best be handled  by employing the \say{Auxiliary  Sweep} feature in the \say{Extended Study} section of \say{Stationary} node. This feature redefines the problem into a sequential solution related to a selection of values, for the load (or displacement), which is taken as the sweep parameter. In the case of \say{Time Dependent} study, the problem is inherently history dependent and the sweep option is not applicable. In order to retain the robustness of the solution during the crack propagation process, the crack increment is kept as small as possible such that the stress redistribution due to generation of new crack surfaces can be handled by the nonlinear Newton-Raphson solver of the software. 

\textbf{Remark 2.} As a result of crack propagation, for both solution strategies, the enrichment zone evolves and new nodes need to be enriched. Subsequently, a series of modifications must be applied to SMenr and SMtip modules, their variables and zones of influence. However, this task is not performed automatically in COMSOL; instead, the initial geometry is adopted throughout the analysis, disregarding any changes in the domain configuration due to crack propagation. In order to render COMSOL to update the geometry as well as the enriched region, the sweep parameter $sp$ is used in the definition of the variables and constraints that alter due to crack evolution, including the displacement constraint that is imposed by \say{Prescribed Displacement} (Eq. \ref{eq:interpolconstraint}) and the modified strain definition (Eq. \ref{eq:epsenrimplement} and Eq. \ref{eq:epstipimplement}) in SMenr and SMtip modules. This is simply achieved by adding a fictitious $\lambda \cdot sp$ to these terms, whereas $\lambda$ is assigned to a very small value (i.e., $\simeq 0$) such that it does not introduce any notable error to the solution.

\textbf{Remark 3.} XFEM modeling of cracks, in essence, is a sequential trinary analysis which consists of pre-processing and level-set update, solution of the governing equations, and post-processing and crack propagation stages.  This requires  that in addition to the pre-processing task that is needed to initialize the problem, certain processes must be executed following each crack increment. This includes retrieving the updated crack details (e.g., crack tip locations and crack body orientations) and corresponding enriched zones. COMSOL does not automatically elaborate such sequencing and therefore, this needs to be effected by the developer. To this end, the \say{Global Variable Probe} tool is used after each step of the analysis to monitor the state of field quantities of the domain, and to store the history variables that are updated in the previous step. Several MATLAB functions are called after each step of the analysis, which include: (i) \say{readcrack.m} and \say{lastangle.m} functions, that provide the previous crack-tip locations and crack increment angles, used in the calculation of the SIFs, and  (ii) \say{crackupdate.m} function, that updates the crack configuration according to the calculated SIFs in conjunction with crack propagation criteria, and modifies the field variable $\psi$ which is used to determine the enriched zone for the next step of the solution. The overall implementation procedure of the XFEM implementation in COMSOL is presented in Algorithm \ref{alg:localmins}.

\begin{algorithm}[t]
\caption{Step by step implementation of XFEM in COMSOL.} 
\label{alg:localmins}
\begin{algorithmic}
\STATE 1. Global Definitions
\STATE \hspace{10mm} Define all constants (material, load, etc)
\STATE \hspace{10mm} Define MATLAB functions (\say{phi.m}, \say{interpol.m}, etc)
\STATE 2. Create Geometry (2D/3D)
\STATE 3. Local Variables definition
\STATE \hspace{10mm} Define interaction integral equations
\STATE \hspace{10mm} Assign Global Variable Probe; call MATLAB functions for crack update, crack tip and angle
\STATE 4. Select physical model (Modules)
\STATE \hspace{10mm} Standard Solid Mechanics (SMstd)
\STATE \hspace{17mm} Select material model
\STATE \hspace{17mm} Select shape function
\STATE \hspace{17mm} Modify stress definitions (Eq. \ref{eq:sigmacomsol})
\STATE \hspace{10mm} Discontinuous enriched Solid Mechanics (SMenr)

\STATE \hspace{17mm} Select material model
\STATE \hspace{17mm} Select shape function
\STATE \hspace{17mm} Modify strain definitions (Eq. \ref{eq:epsenrimplement})
\STATE \hspace{17mm} Modify stress definitions (Eq. \ref{eq:sigmacomsol})
\STATE \hspace{17mm} Apply field variable $\psi$ as a constraint using domain Prescribed Displacement option

\STATE \hspace{10mm} Crack tip enriched Solid Mechanics (SMtip)

\STATE \hspace{17mm} Select material model
\STATE \hspace{17mm} Select shape function
\STATE \hspace{17mm} Modify strain definitions (Eq. \ref{eq:epstipimplement})
\STATE \hspace{17mm} Modify stress definitions (Eq. \ref{eq:sigmacomsol})

\STATE 5. Assign initial and boundary conditions
\STATE 6. Discretization and mesh generation
\STATE 7. Specify Study type
\STATE \hspace{10mm} Select Parametric Sweep analysis
\STATE 8. Post-processing and visualization

\end{algorithmic}

\end{algorithm}

%% file: results.tex
 \section{Numerical simulations}
\label{S:4 (results)}

In this section, the accuracy and robustness of the proposed XFEM implementation in COMSOL are thoroughly investigated by several numerical simulations.  In the first example, the performance of the proposed solution strategy is investigated in the case of stationary cracks. A convergence study is conducted and the SIF values of an inclined crack for pure mode I and mixed-mode cases are acquired, and compared to the available analytical solutions in the literature. The flexibility of the proposed implementation to handle heavily fractured domains is demonstrated in another 2D example. In the subsequent two examples, mixed-mode crack propagation in complex geometries is studied comprehensively. Finally, a selection of three-dimensional fracture analysis is carried out to illustrate the capability of the proposed implementation in dealing with more complex geometric settings. 

In all the examples, a linear elastic material with Young's modulus of $E=200\text{GPa}$ and Poisson's ratio of $\nu=0.3$ is supposed, unless specified otherwise. Quasi-static formulation is used for crack propagation analysis under displacement controlled boundary conditions. The 2D analyses are performed by assuming plane strain state, and bi-linear quadrilateral elements are used to discretize the solution domain. Tetrahedral and brick elements are employed for 3D analysis. The integration in the enriched modules SMenr and SMtip is carried out by using 35-point and 40-point Gaussian quadrature, respectively, while $\delta$ is set to 0.002 to ensure the existence of sufficient number of Gauss integration points at either sides of the crack interfaces, within the enriched elements.

\subsection{Center crack in an infinite domain; model verification and SIF analysis}
\label{S:4-1 (middlecrackSIF)}

\begin{figure}[!t]
\centering\includegraphics[width=0.45\linewidth]{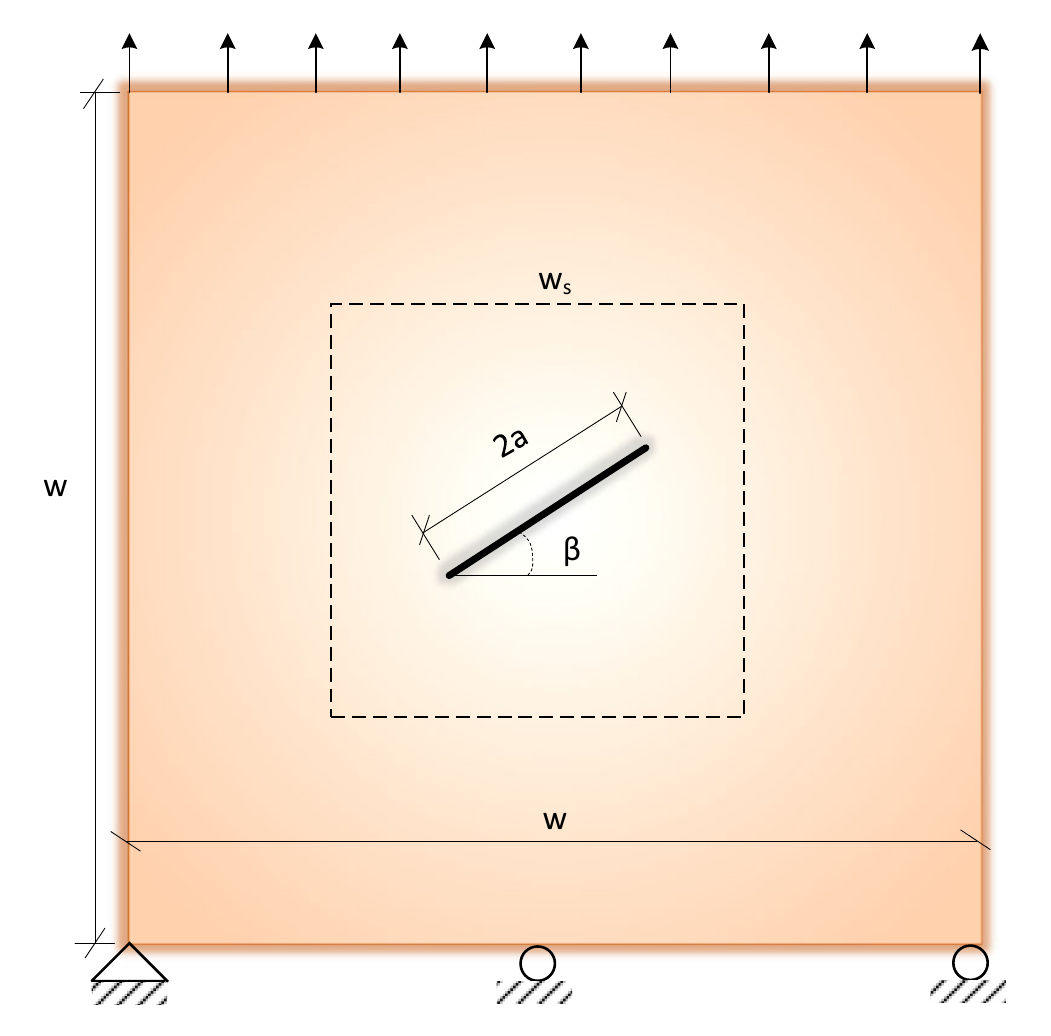}
\caption{Geometry and boundary conditions of a center crack in an infinite domain.}
\label{fig:ex1-SIF}
\end{figure}

In this example, the simulation results associated with the proposed XFEM implementation are compared to a series of benchmark analytic solutions in 2D settings \cite{anderson2017fracture}. As depicted in Fig. \ref{fig:ex1-SIF}, a square plate is considered with side length of $w=5\text{ m}$ that contains an inclined center crack of size $2a=0.2\text{ m}$.  The ratio $w/a$ is chosen as $50$ to emulate a crack in an infinite domain. The plate is subjected to uniaxial far-field tension of $1$ MPa at the top edge, while the bottom edge is fixed. 

In the first part of this example, a convergence study is performed on the stress intensity factors and the significance of the crack tip enrichment for the case of a horizontal center crack ($\beta=0$). To this end, the mesh in the vicinity of each crack-tip (i.e., over a square of length $w_{\text{s}}=1\text{m}$) is refined using the normalized element sizes of $a/s = 2,5,6.7,9.1$ and $12.5$, where $s$ is the element size in the refinement zone. The radius of the circular integration path employed for the interaction-integral calculations is set to $a$ (see section \ref{S:3-2 module}). In Table \ref{t:meshsens},  the simulations results of the developed model are compared against the exact solutions expressed as,

\begin{equation}
\label{eq:sifanal}
\begin{matrix}
K_{\text{I}}=\sigma \sqrt{\pi a}\text{ cos}^{2}\beta \\ 
K_{\text{II}}=\sigma \sqrt{\pi a}\text{ sin}\beta \text{ cos}\beta 
\end{matrix}
\end{equation}

\begin{table}[ht]
\caption{Relative errors of the stress intensity factors for a horizontal crack in infinite plate with (w) and without (w/o) crack tip enrichment (SIFs are in MPa$\sqrt{\text{m}}$).}
\centering 
\begin{tabular}{c c c c c c c}
\hline \hline 
$\bar{a}^{*}(\text{m})$ & $a/s$ & $K_{\text{I}}^\text{exact}$  & $K_{\text{I}}^{\text{w/o tip}^{\textcolor{white}{A}}}$  & $K_{\text{I}}^{\text{w tip}^{\textcolor{white}{A}}}$   & $\text{error}^{\text{w/o tip}}(\%)$ & $\text{error}^{\text{w tip}}(\%)$ \\ [0.5ex]
\hline                  
0.1 &	2	    & 0.5605 & 0.5893 & 0.5658 &	4.88 & 0.94 \\
0.1	&   5       & 0.5605 & 0.5731 & 0.5632 & 2.20   & 0.48 \\
0.108	& 6.7	& 0.5825 & 0.5742 & 0.5588 & 1.42  & 0.29\\
0.103	& 9.1	& 0.5675 & 0.5643 & 0.5601 & 0.55 & 0.07\\
0.102	& 12.5	& 0.5647 & 0.5650 & 0.5605 & 0.05 & 0.00\\ [1ex]      
\hline
\end{tabular}\\
      \small
       * $\bar{a}$ represents the crack length in the COMSOL model w/o tip enrichment; this can be slightly different from the nominal crack length, since the enriched elements are considered fully fractured up to the element edges.
\label{t:meshsens}
\end{table}
As can be seen from Table \ref{t:meshsens}, the proposed procedure evaluates the SIFs correctly and with high accuracy. It is also observed that Employing the crack tip enrichment functions in the model can minimize the errors in SIFs even for relatively coarse discretizations; however, this is achieved at the expense of increased computational cost, since, four additional Solid Mechanics modules with maximized integration order are required to accommodate the asymptotic crack tip functions. On the other hand, the results for the cases where the crack tip enrichment functions are excluded show a satisfying accuracy for the range of $a/s>7$. Hence, in favour of computational efficiency and simplicity in implementation, from here onward merely the discontinuous Heaviside enrichment is considered in the following examples.

\begin{figure}
\centering
\begin{tikzpicture}
\begin{axis}[
    xlabel={ Log (mesh size (m))},
    ylabel={Log ($\left \| e \right \|_{E}$)},
    xmin=-2.2, xmax=-0.7,
    ymin=-2.2, ymax=-0.7,
    xtick={-2,-1.5,-1,-0.5},
    ytick={-2,-1.5,-1,-0.5},
    legend pos=south east,
    ymajorgrids=false,
    grid style=dash,
]

\addplot[
    color=blue,
    dashed,
    mark=non,
    ]
    coordinates {
    (-2,-1.8681)(-1,-0.9094)
    };
    
\addplot[
    only marks,
    color=blue,
    mark=square,
    ]
    coordinates {
    (-2,-1.843875053)(-1.602059991, -1.528855035)(-1.301029996, 	-1.194274916
    )(-1, -0.895273956)};
   \addlegendentry{$m=0.96$ ($R^{2}=0.99$)}

\end{axis}
\end{tikzpicture}
\caption{Energy error norm for the horizontal crack problem using the proposed XFEM implementation.}
\label{fig:errornorm}
\end{figure}
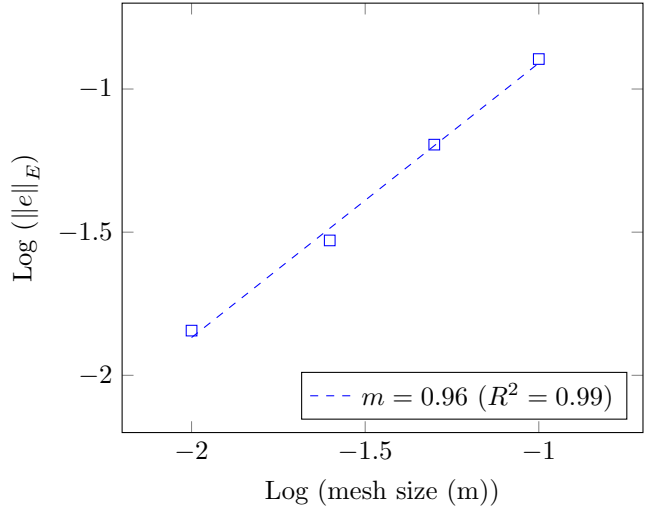

In Fig. \ref{fig:errornorm}, the convergence in the energy error norm of the proposed formulation versus element size is studied. The energy error norm $\left \| e \right \|_{E}$ is defined as \cite{liu2013smoothed}

\begin{equation}
\label{eq:errornorm}
\left \| e \right \|_{E}=\frac{1}{\Omega }\sqrt{\int_{\Omega }^{}(\mathbf{\varepsilon}^{\text{XFEM}}-\mathbf{\varepsilon}^{\text{exact}})\mathbf{D}(\mathbf{\varepsilon}^{\text{XFEM}}-\mathbf{\varepsilon}^{\text{exact}})d\Omega  }
\end{equation}
where $\mathbf{\varepsilon}^{\text{XFEM}}$ is the strain field associated with the XFEM simulation, whereas $\mathbf{\varepsilon}^{\text{exact}}$ is the high-fidelity solution due to a FEM analysis using an extremely fine mesh. The rate of variations in the error norm is used to demonstrate the validity of the numerical analysis. Fig. \ref{fig:errornorm} demonstrates that the optimal convergence rate of almost 1 is achieved by the proposed XFEM approach \cite{khoei2014extended}. 

\begin{figure}
\centering
\begin{tikzpicture}
\begin{axis}[
    xlabel={ $\beta$ (deg)},
    ylabel={$K_{\text{I}}$,$K_{\text{II}}$ (MPa$\sqrt{\text{m}}$)},
    xmin=0, xmax=90,
    ymin=0, ymax=0.6,
    xtick={0,10,20,30,40,50,60,70,80,90},
    ytick={0.0,0.1,0.2,0.3,0.4,0.5,0.6},
    legend pos=north east,
    ymajorgrids=false,
    grid style=dash,
]

\addplot[
    color=blue,
    mark=square,
    ]
    coordinates {
    (10,0.546)(20,0.498)(30,0.422)(40,0.331)(50,0.233)(60,0.141)(70,0.066)(80,0.017)
    };
   \addlegendentry{$K_{\text{I}}$ COMSOL}

\addplot[
    color=blue,
    mark=triangle,
    ]
    coordinates {
    (10,0.094)(20,0.184)(30,0.251)(40,0.286)(50,0.284)(60,0.251)(70,0.184)(80,0.095)
    };
    \addlegendentry{$K_{\text{II}}$ COMSOL}

\addplot[
    color=red,
    dashed,
    mark=square,
    ]
    coordinates {
    (10,0.544)(20,0.495)(30,0.420)(40,0.329)(50,0.232)(60,0.140)(70,0.066)(80,0.017)
    };
    \addlegendentry{$K_{\text{I}}$ Analytical}    
    
\addplot[
    color=red,
    dashed,
    mark=triangle,
    ]
    coordinates {
    (10,0.096)(20,0.180)(30,0.243)(40,0.276)(50,0.276)(60,0.243)(70,0.180)(80,0.096)
    };
    \addlegendentry{$K_{\text{II}}$ Analytical}
    
\end{axis}
\end{tikzpicture}
\caption{Comparison of $K_{\text{I}}$ and $K_{\text{II}}$ values for the inclined crack problem; COMSOL results vs analytical exact solutions.}
\label{fig:sifcompare}
\end{figure}
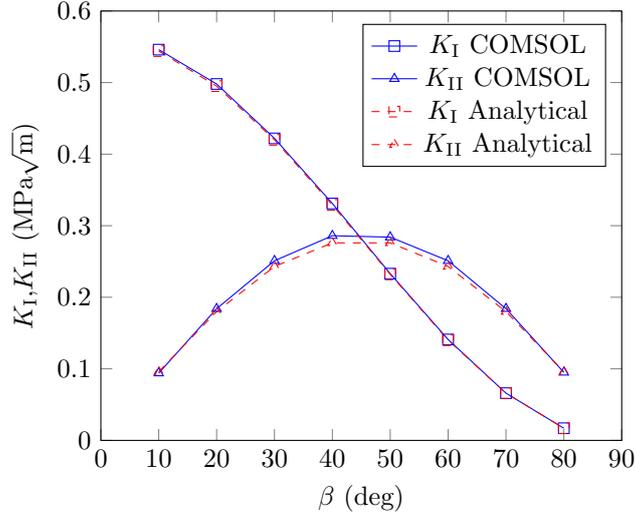

In the remainder of this example, the effectiveness of the proposed procedure in handling mixed-mode fracturing is demonstrated through examining the stress intensity factors for the case of inclined cracks. In this case, values for $K_{\text{I}}$ and $K_{\text{II}}$ are obtained by means of a locally refined mesh, using a normalized element size of $a/s=12$, over a square zone of length $w_{\text{s}}$ that encompasses the crack. A circular path with a radius of $0.9a$ is adopted for the calculation of the interaction integral. All other assumptions are similar to the horizontal crack problem definition. The calculated SIFs for the mixed-mode crack analysis are depicted in Fig. \ref{fig:sifcompare}, which are in excellent agreement with the exact values given by Eq. \ref{eq:sifanal}. Notably, the maximum error in the calculation of the crack propagation angle by means of the obtained SIFs does not exceed $0.5$ degrees (see Eq. \ref{eq:thetac}), which further highlights the accuracy of the proposed approach.

\subsection{Square plate with multiple randomly distributed cracks}
\label{S:4-2 heavilyfractured}

\begin{figure}[!t]
\centering\includegraphics[width=0.45\linewidth]{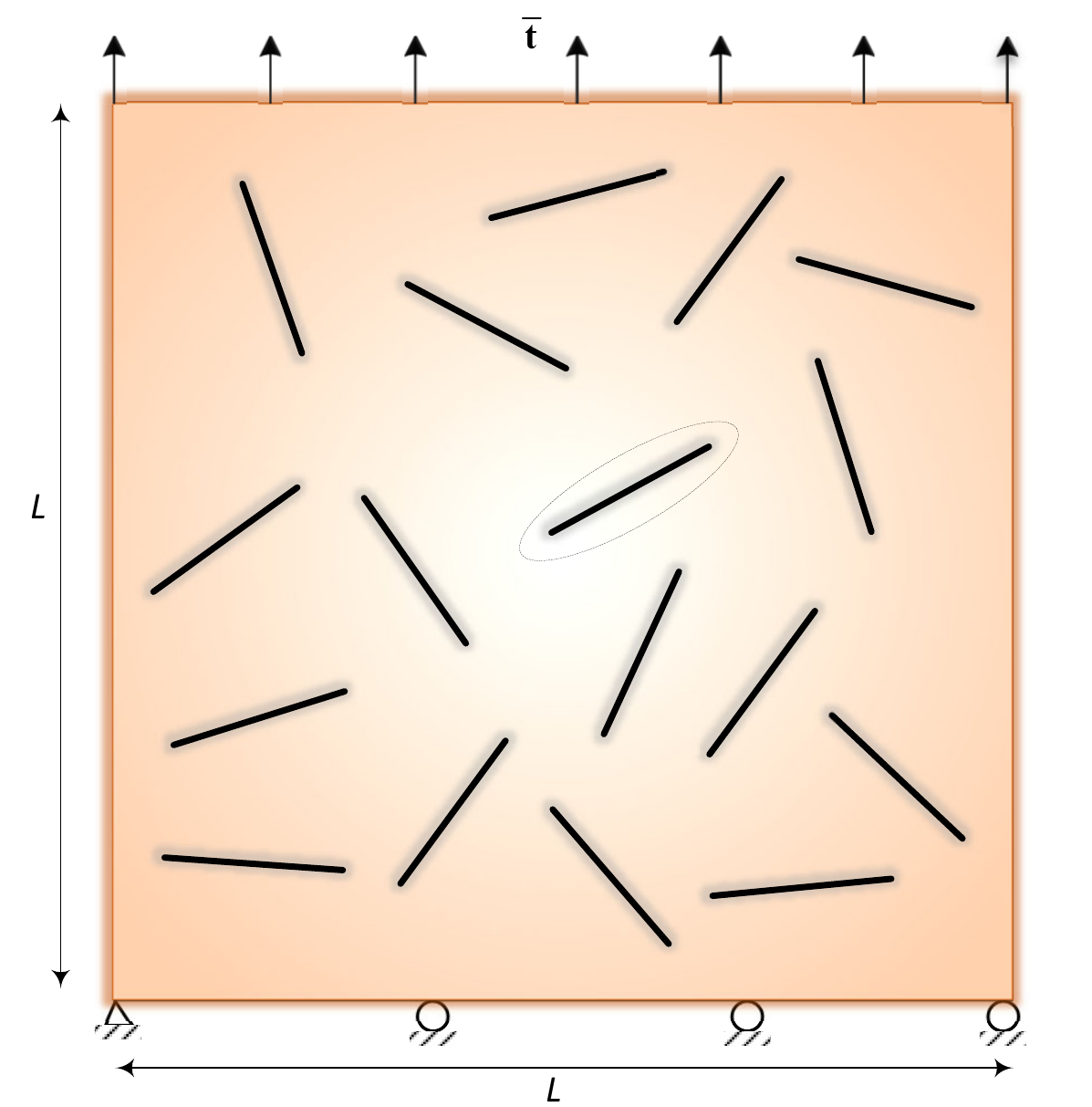}
\caption{Randomly distributed cracks problem; geometry and boundary conditions. The marked crack is chosen to study the crack opening displacement. }
\label{fig:ex-3 geometry}
\end{figure}

The heterogeneity caused due to the presence of pre-existing cracks is a crucial subject in a wide range of research fields including: the micro-mechanical behaviour of concrete \cite{kurumatani2019simulations}, micro-cracks in biological organs \cite{hammond2019mechanics}, and natural fractures in geological formations \cite{vahab2021numerical,hirmand2019robust}, to name a few. This example studies the robustness and flexibility of the proposed implementation in handling domains containing randomly distributed cracks. As Fig. \ref{fig:ex-3 geometry} shows, the problem consists of 17 equally-sized cracks, with the length of $0.2$ m, which are randomly distributed in a square plate of side length $L=1$ m. The plate is subjected to tensile traction of $\bar{\mathbf{t}}=1$ MPa at the top edge, while the bottom edge is supposed to be fixed. The material properties are identical to example \ref{S:4-1 (middlecrackSIF)}. In lieu of exact solution, a high-fidelity FEM model using the same configuration is employed. The domain meshes consists of 14,641 and 15,000 quadrilateral elements, respectively, for the XFEM implementation and the FEM model, in which an average element size of $8$ mm is adopted.

\begin{figure}
\centering
\begin{subfigure}{.5\textwidth}
  \centering
  \includegraphics[width=.95\linewidth]{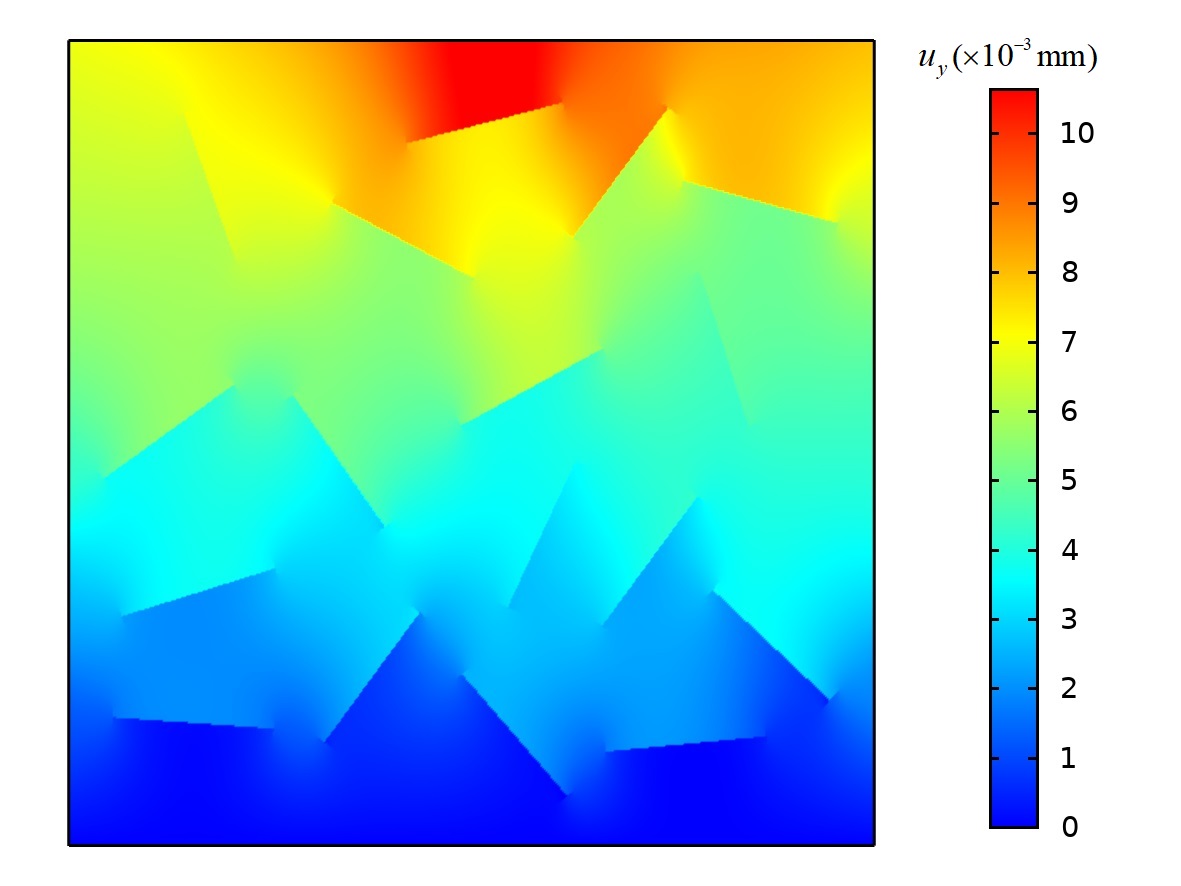}
  \caption{XFEM}
  \label{fig:ex3-UyXFEM}
\end{subfigure}%
\begin{subfigure}{.5\textwidth}
  \centering
  \includegraphics[width=0.95\linewidth]{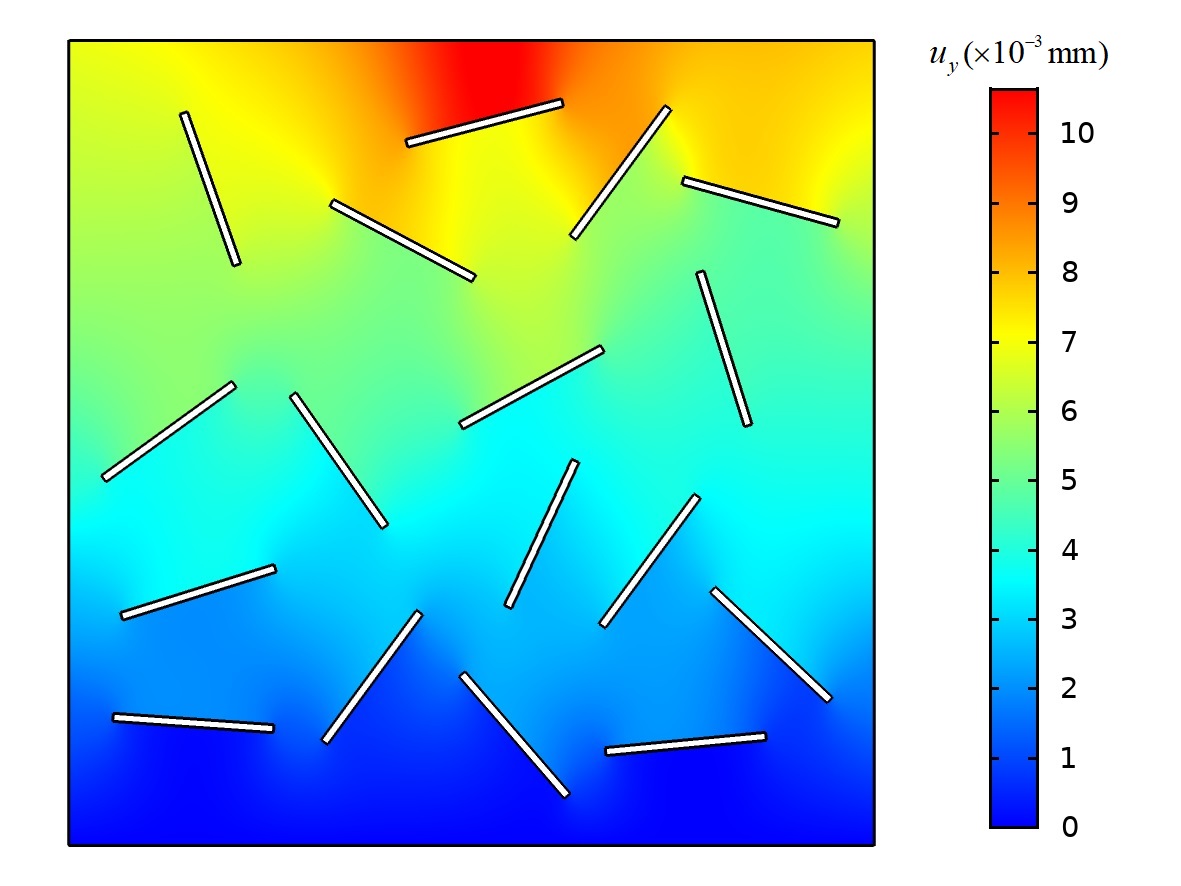}
  \caption{FEM}
  \label{fig:ex3-UyFEM}
\end{subfigure}
\caption{Contours of vertical displacement distribution $u_{\text{y}}$ in the domain containing randomly distributed cracks; XFEM vs FEM results.}
\label{fig:ex3-Uy}
\end{figure}

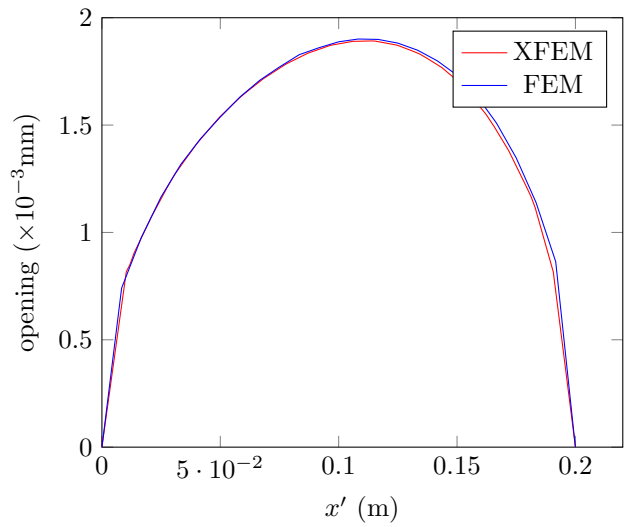
\begin{figure}
\centering
\begin{tikzpicture}
\begin{axis}[
    xlabel={ ${x}'$ (m)},
    ylabel={opening ($\times10^{-3}$mm)},
    xmin=0, xmax=0.22,
    ymin=0.0, ymax=2,
    xtick={0,0.05,0.10,0.15,0.2},
    ytick={0.0,0.5,1.0,1.5,2.0},
    legend pos=north east,
    ymajorgrids=false,
    grid style=dash,
]

\addplot[
    color=red,
    mark=false,
]
    file[] {openingXFEM.dat};
   \addlegendentry{XFEM}

\addplot[
    color=blue,
    mark=false,
]  
 file[]{openingFEM.dat};
    \addlegendentry{FEM}

\end{axis}
\end{tikzpicture}
\caption{Crack opening displacement profile of an arbitrary crack in a domain with randomly distributed cracks; XFEM vs FEM results.}
\label{fig:ex3opening2D}
\end{figure}

\begin{figure}
\centering
\begin{subfigure}{.5\textwidth}
  \centering
  \includegraphics[width=.95\linewidth]{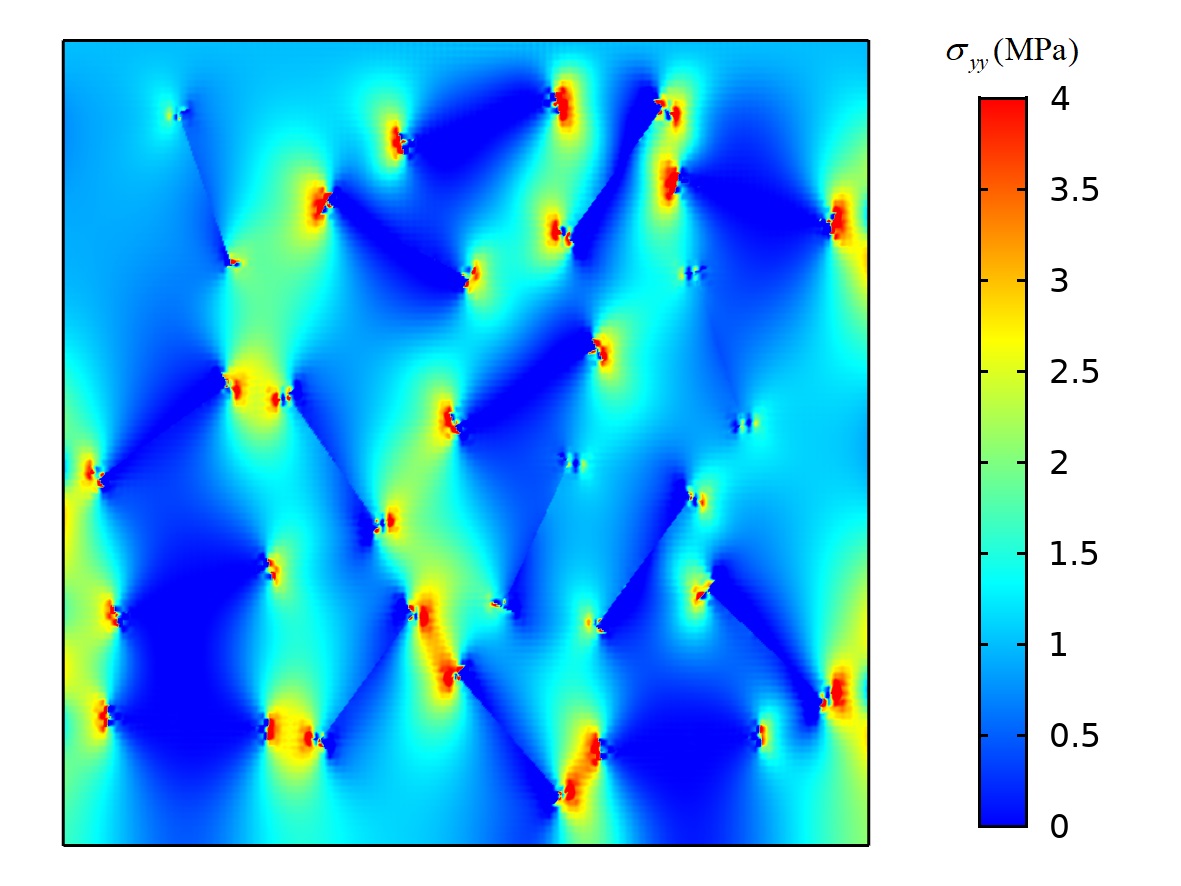}
  \caption{XFEM}
  \label{fig:ex3-SyyXFEM}
\end{subfigure}%
\begin{subfigure}{.5\textwidth}
  \centering
  \includegraphics[width=0.95\linewidth]{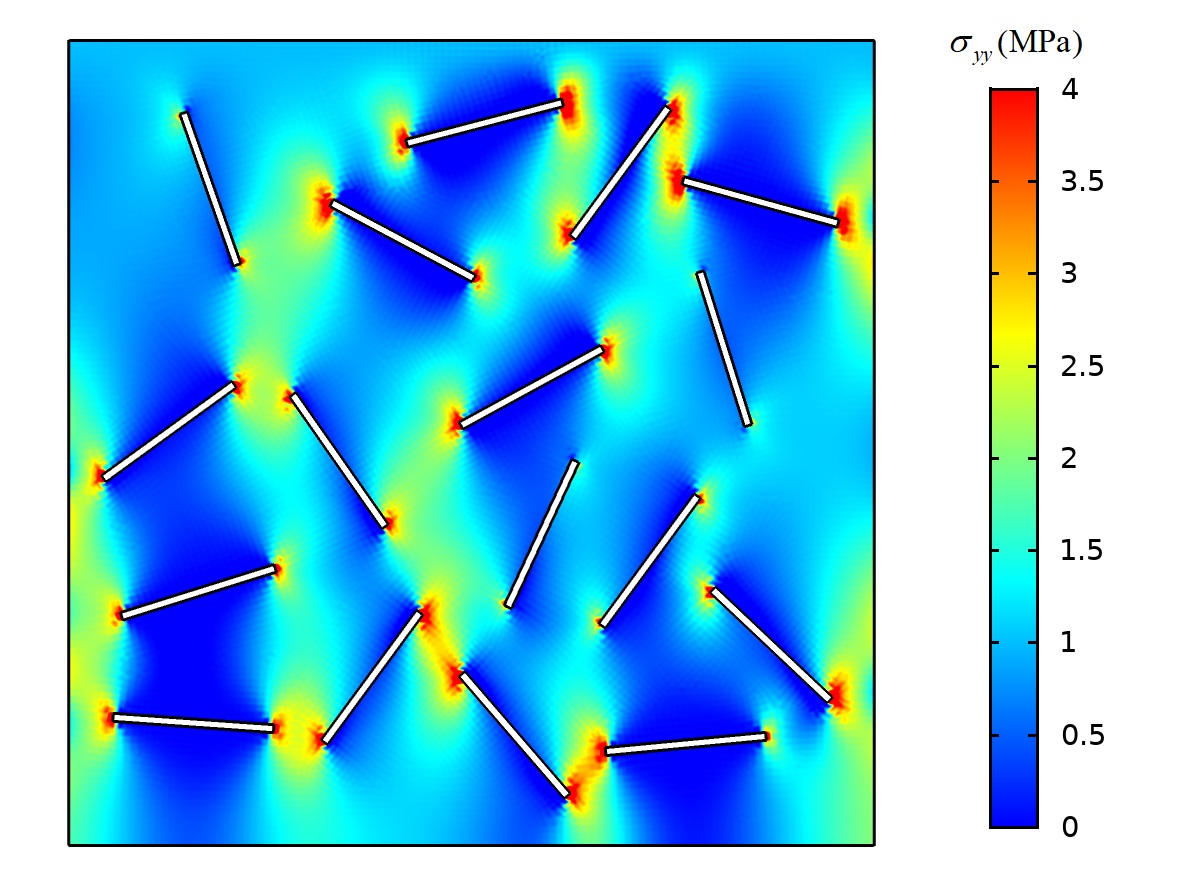}
  \caption{FEM}
  \label{fig:ex3-SyyFEM}
\end{subfigure}
\caption{Contours of vertical stresses $\sigma _{\text{yy}}$ in a domain with randomly distributed cracks.}
\label{fig:ex3=Syy}
\end{figure}

Fig. \ref{fig:ex3-Uy} shows the contours of the vertical displacement $u_y$ for both XFEM and FEM simulations, where an excellent agreement is observed between the results. In addition, the profile of crack opening displacement (COD) is presented for one of the cracks in Fig. \ref{fig:ex3opening2D} (marked by an ellipse in Fig. \ref{fig:ex-3 geometry}) . It is observed that the maximum difference between the two opening profiles is less than $2\%$. Note minor discrepancies in here are attributed to the introduction of crack surfaces by finite width bodies of width $0.01$ m in FEM model. This is accompanied by the need for extraction of fracture profile by means of displacement field at either sides of the crack in FE model; yet, this task in the XFEM implementation is readily performed by using the enriched component of the displacement field (i.e., $\mathbf{u}^{\text{disc}}$ in SMenr). Finally, contours of the vertical stress $\sigma_{yy}$ is presented in Fig. \ref{fig:ex3=Syy} for the both simulations. Both contours match reasonably well, in particular, at locations the cracks-tips stress fields are interrupted by each other. 

\subsection{Mixed-mode crack propagation}
\label{S:4-2 (crackpropagate)}

The following two numerical examples are presented to show the outstanding applicability of the proposed implementation in dealing with mixed-mode crack propagation in complex geometries. In both cases, quasi-static loading condition is considered, while the propagation angle of the cracks is determined based on the calculated SIFs during the course of the analysis (see Eq. \ref{eq:thetac}).

\subsubsection{Crack propagation in a rectangular plate with a hole}
\label{S:4-2-1 crackhole}

\begin{figure}
\centering
  \includegraphics[width=.35\linewidth]{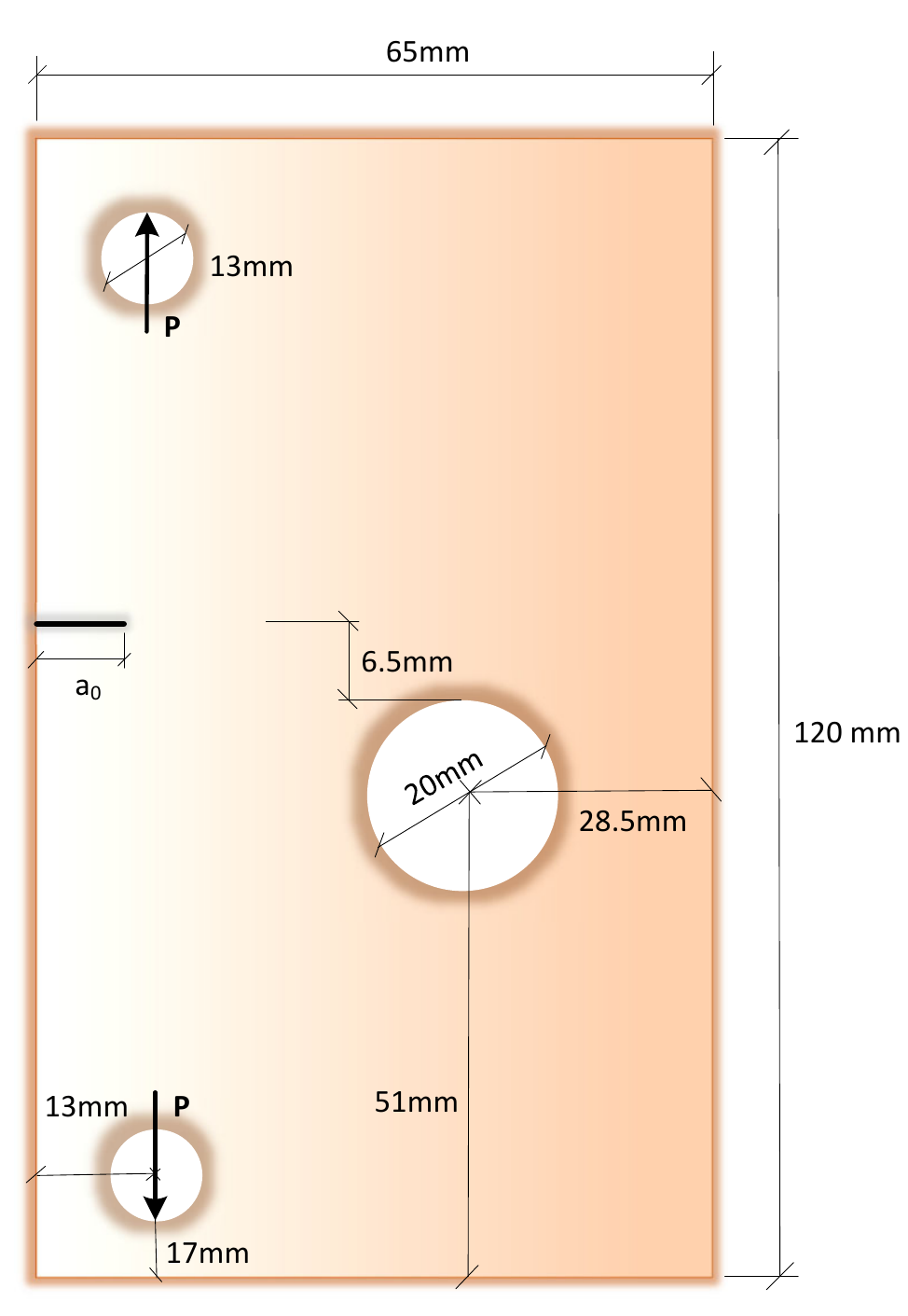}
\caption{Crack in a plate with hole: Geometry and boundary conditions.}
\label{fig:ex2-1geo}
\end{figure}

This example is adopted from Giner et. al. \cite{giner2009abaqus}, which aims to investigate the effects of a hole in a rectangular plate on the crack propagation pattern.  Fig. \ref{fig:ex2-1geo} illustrates the geometry and boundary conditions of the plate, that is made of an aluminum alloy with $E=71.7\text{ GPa}$ and $\nu=0.33$. Consistent with the reference, the initial crack length and crack growth increment are set to $a_{0}=10\text{ mm}$ and $\Delta a=3\text{ mm}$, respectively. An incrementally increasing traction is applied to the top edge of the plate with a maximum of 15 kN/m. The domain is discretized with 7,601 quadrilateral elements with an average element size of $0.67\text{ mm}$. 

\begin{figure}
\centering
\begin{subfigure}{.33\textwidth}
  \centering
  \includegraphics[width=.70\linewidth]{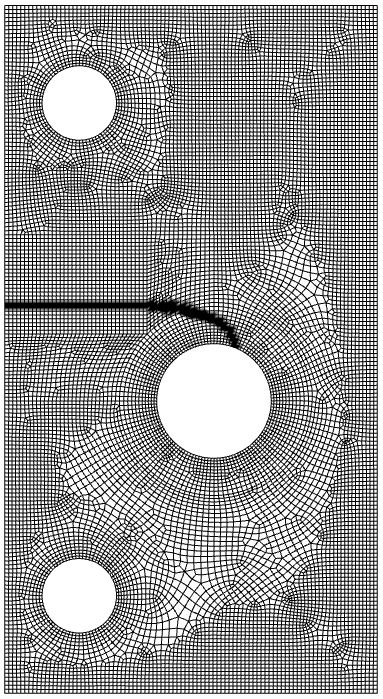}
  \caption{}
  \label{fig:ex2-1path}
\end{subfigure}%
\begin{subfigure}{.33\textwidth}
  \centering
  \includegraphics[width=0.90\linewidth]{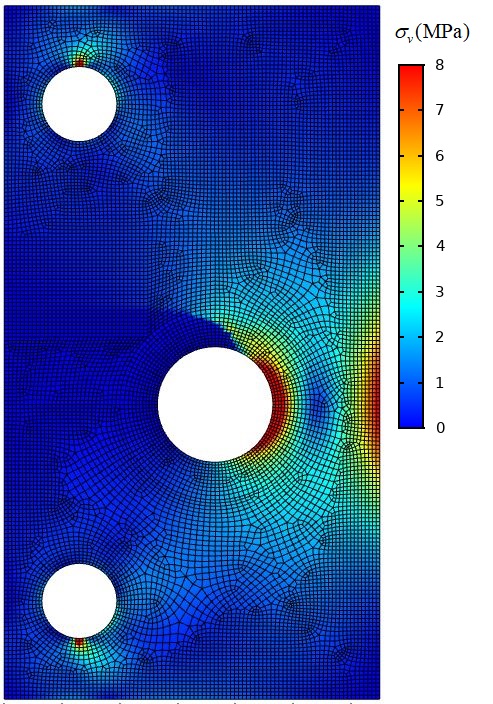}
  \caption{}
  \label{fig:ex2-1mises}
\end{subfigure}
\begin{subfigure}{.33\textwidth}
  \centering
  \includegraphics[width=.8\linewidth]{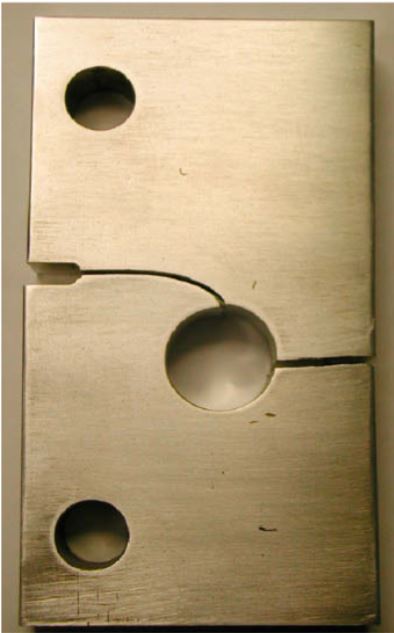}
  \caption{}
  \label{fig:ex2-1exp}
\end{subfigure}
\caption{Crack propagation in a rectangular plate with hole; a) crack trajectory based on $\psi$ field, b) von-Mises stress $\sigma_{\text{v}}$ contour at the end of the analysis, and c) experimental observations by \cite{giner2009abaqus}.}
\label{fig:ex2-1pathmises}
\end{figure}

Figs. \ref{fig:ex2-1path} and \ref{fig:ex2-1mises} respectively show the crack trajectory and the von-Mises stress contour at the end of the analysis. The former is deduced by using the field variable $\psi$, which is equal to unity for fractured elements. The numerical results are in excellent agreement with the experimental observations as depicted in Fig. \ref{fig:ex2-1exp}.

\subsubsection{Multiple Crack propagation in a plate with double holes}
\label{S:4-2-2 crackhole}

\begin{figure}[!t]
\centering\includegraphics[width=0.7\linewidth]{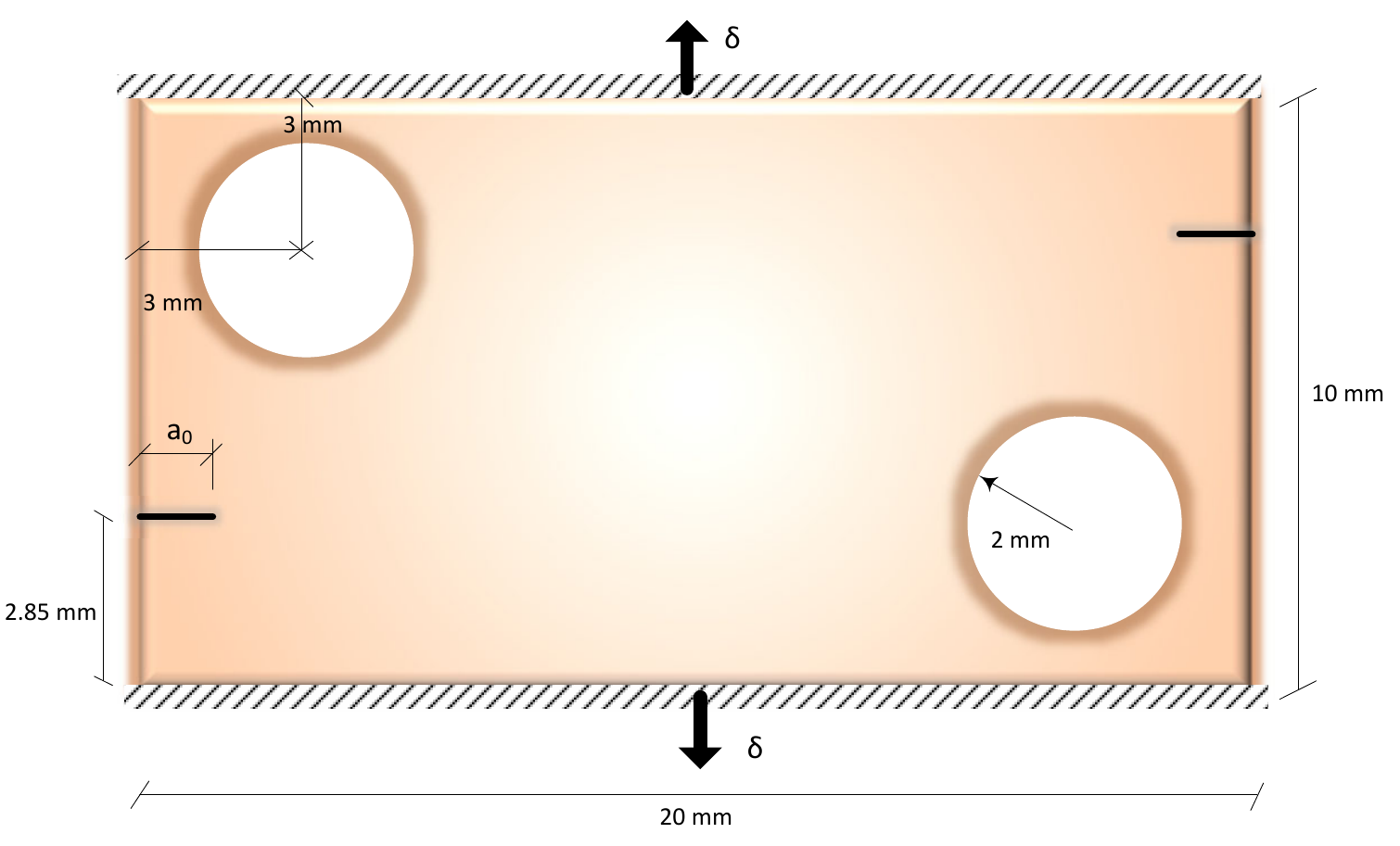}
\caption{A plate with two holes and multiple cracks; problem geometry and boundary conditions.}
\label{fig:ex2-2geomtry}
\end{figure}

\begin{figure}
\centering
\begin{subfigure}{.5\textwidth}
  \centering
  \includegraphics[width=.93\linewidth]{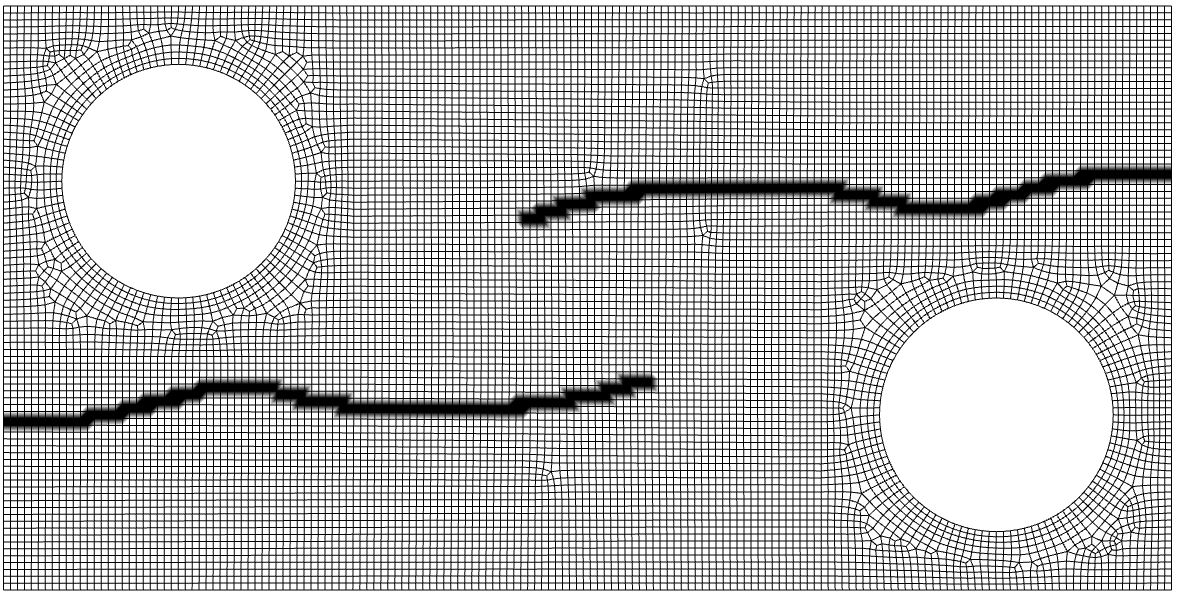}
  \caption{}
  \label{fig:ex2-2pathcomsol}
\end{subfigure}%
\begin{subfigure}{.5\textwidth}
  \centering
  \includegraphics[width=0.95\linewidth]{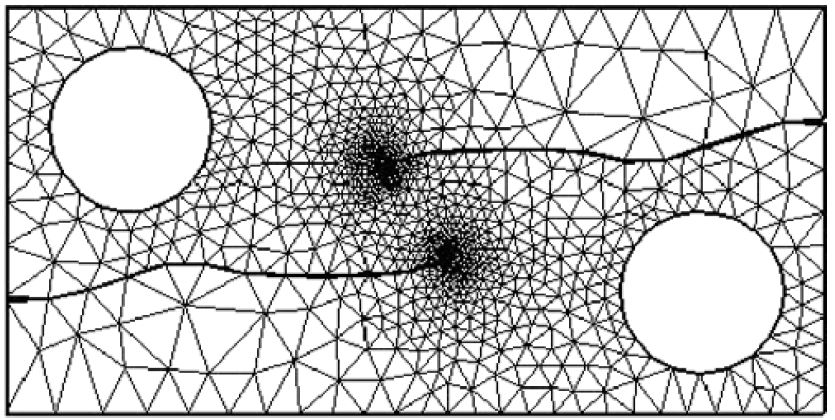}
  \caption{}
  \label{fig:ex2-2pathadaptive}
\end{subfigure}
\caption{Crack propagation trajectory in a plate with two holes: a) COMSOL simulation and b) adaptive finite element method \cite{khoei2008modeling}.}
\label{fig:ex2-2path}
\end{figure}

\begin{figure}[!t]
\centering\includegraphics[width=0.55\linewidth]{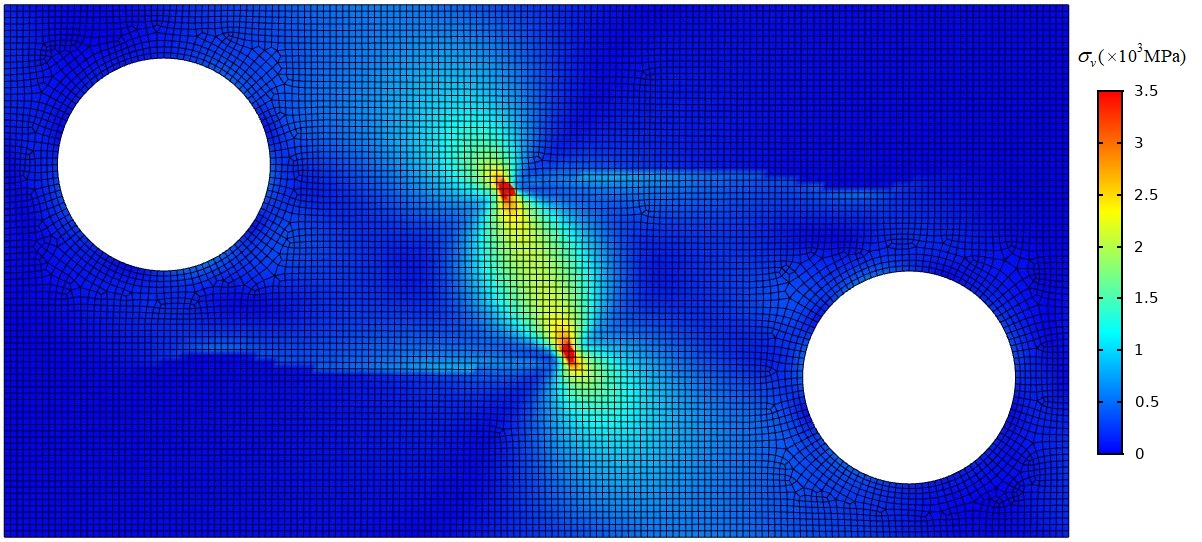}
\caption{Final von-Mises stress contour in a plate with two holes and multiple cracks.}
\label{fig:ex2-2mises}
\end{figure}

In the second example, mixed-mode crack propagation in a plate involving two holes is investigated by means of the proposed model (Fig. \ref{fig:ex2-2geomtry}). Note this problem was originally introduced by Bouchard et. al. \cite{bouchard2003numerical}. The aim of the simulation is to further demonstrate the capability of the proposed implementation in dealing with multiple crack propagation in more complex geometries. Due to the ideal antisymmetry incorporated in definition of the geometry, FE mesh and boundary conditions, both of the pre-cracks are expected to propagate identically. The initial length of both cracks is $a_{0}=1\text{mm}$, and the critical fracture toughness is taken as $K_{\text{IC}}=47.4\text{ MPa}\sqrt{\text{m}}$. Two sets of variables, corresponding to each crack tip, are introduced in order to calculate and store the SIFs during the solution. To retain the antisymmetry of the solution, the plate is subjected to prescribed vertical displacement $\delta=0.05\text{ mm}$ at both top and bottom edges. The simulation is performed by means of a FE mesh with 12,149 quadrilateral elements. To ensure best outcome, the mesh is designed to be relatively structured and symmetrical. As Fig. \ref{fig:ex2-2pathcomsol} shows, the cracks initially deviate towards the adjacent holes and then gradually realign with their initial path. This perfectly matches the crack trajectory obtained by Khoei et al. \cite{khoei2008modeling} using the adaptive finite element method, as depicted in Fig. \ref{fig:ex2-2pathadaptive}. The corresponding von-Mises stress distribution contour at the final crack increment is presented in Fig. \ref{fig:ex2-2mises}.

\subsection{Penny-shaped crack in three-dimensional media}
\label{S:4-4 planar3D}

\begin{figure}[!t]
\centering\includegraphics[width=0.5\linewidth]{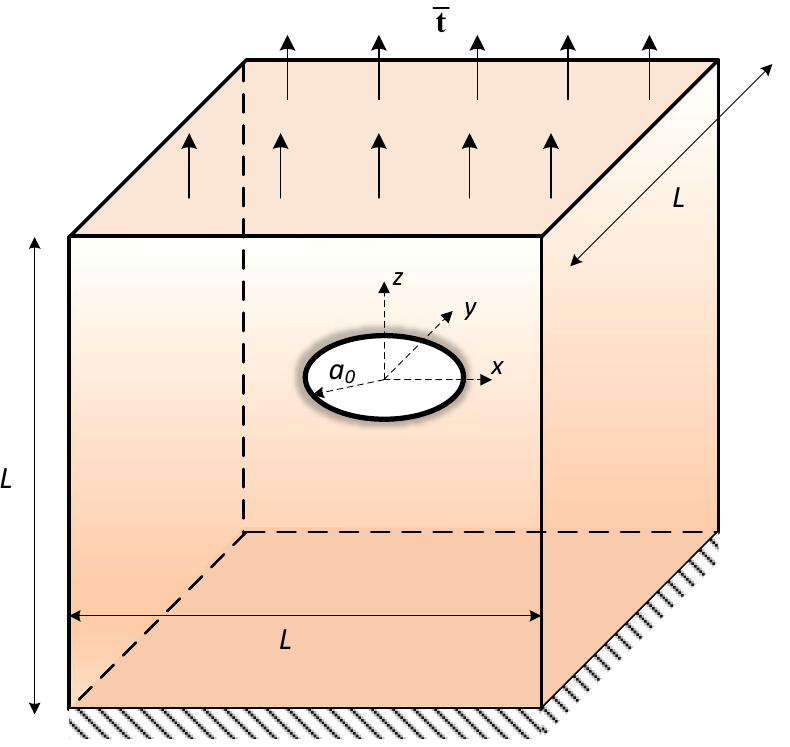}
\caption{A cube with penny-shaped crack problem; geometry and boundary conditions.}
\label{fig:ex-4 geometry}
\end{figure}

The final example is presented to illustrate the extendibility of the proposed XFEM implementation to three dimensional problems. As depicted in Fig. \ref{fig:ex-4 geometry}, the problem consists of a cube with side length of $1$ m which contains a penny-shaped crack of radius $0.25$ m at its center. The top surface is subjected to a tensile traction of $10$ MPa, and the bottom surface is fixed. The domain is discretized by a cluster of 50,280 brick elements with average size of $13$ mm in the vicinity of the crack zone, in conjunction with additional 100,843 tetrahedral elements in the remainder of the domain. For the sake of simplicity, here the penny-shaped geometry is explicitly introduced in COMSOL by mathematical relations, instead of the more general yet complicated procedure through MATLAB functions (i.e., \say{phi.m} and \say{interpol.m}). For comparison purposes, an FEM analysis of the same problem is performed in which the penny-shaped crack is modeled by a narrow cylindrical void.

\begin{figure}
\centering
\begin{subfigure}{.5\textwidth}
  \centering
  \includegraphics[width=.95\linewidth]{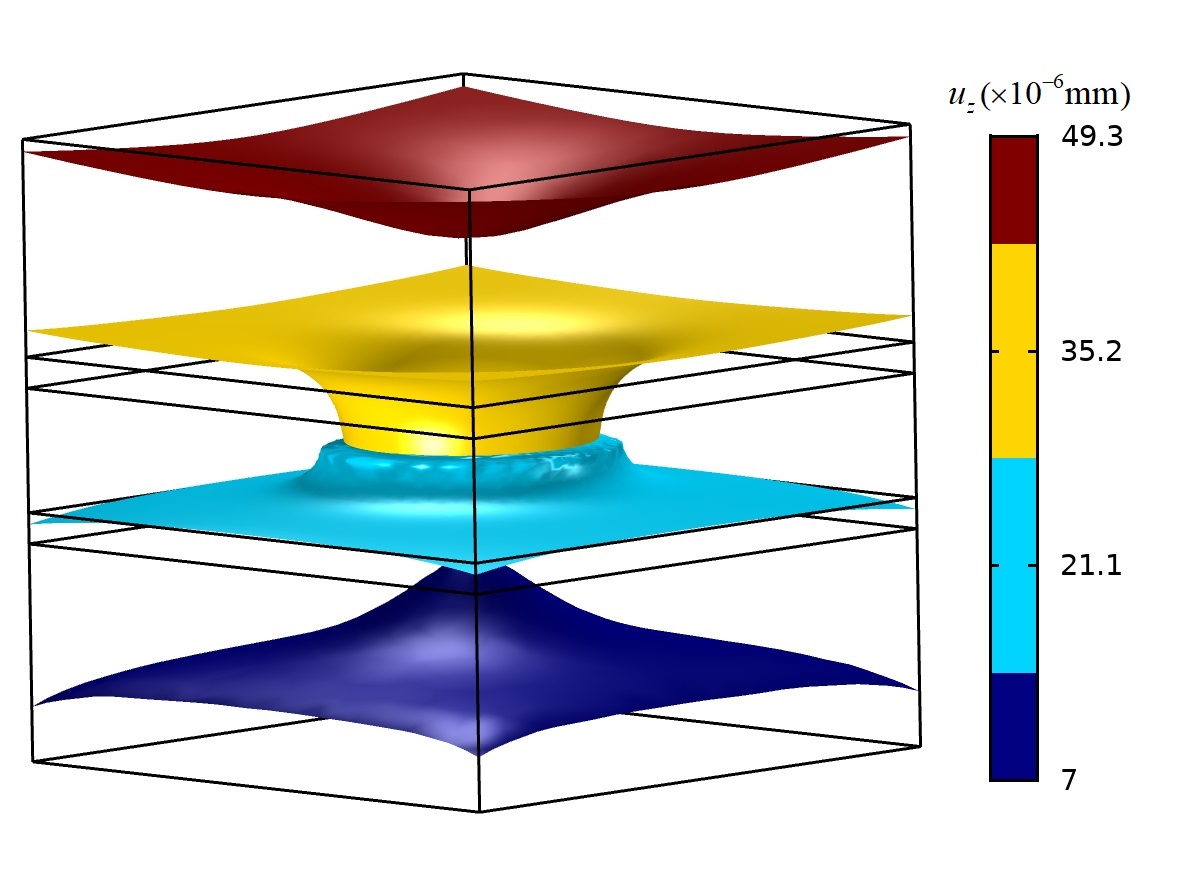}
  \caption{XFEM}
  \label{fig:ex4-isodurfaceXFEM}
\end{subfigure}%
\begin{subfigure}{.5\textwidth}
  \centering
  \includegraphics[width=0.95\linewidth]{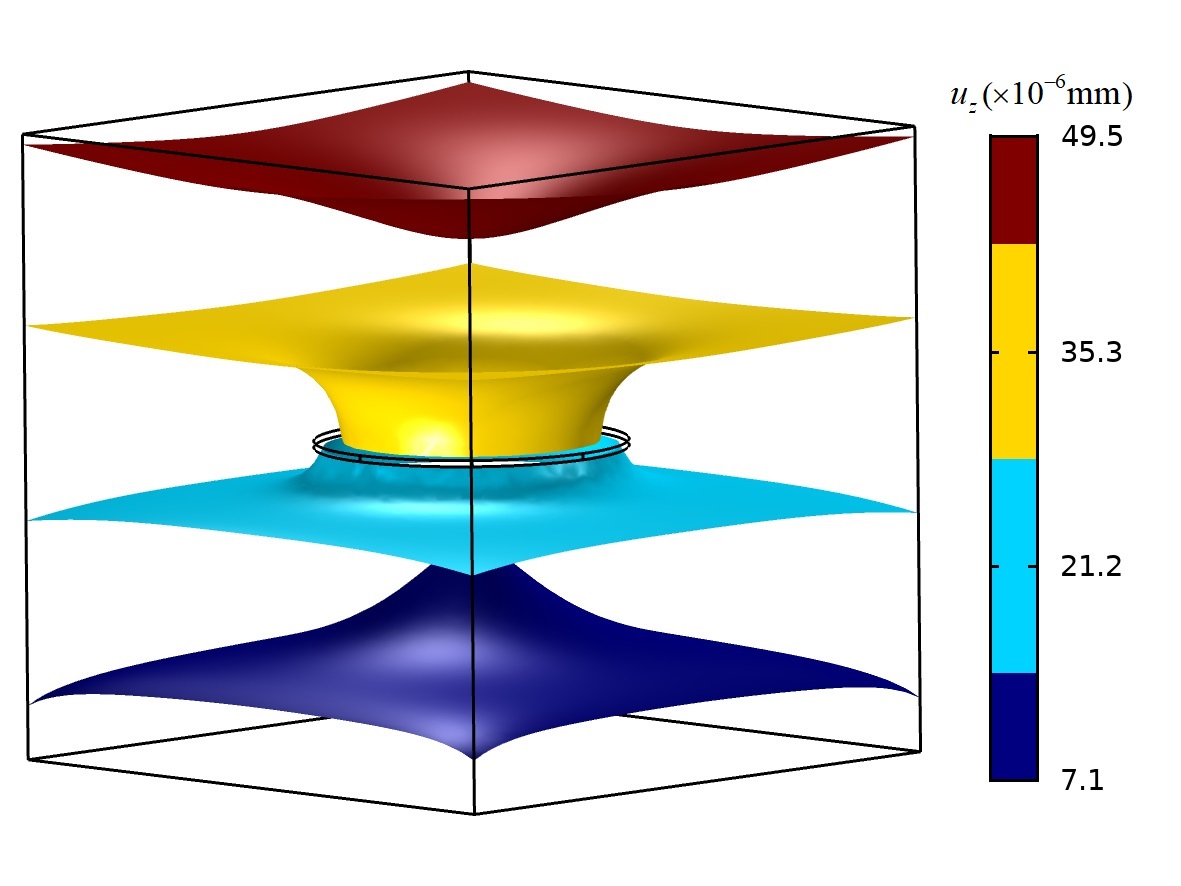}
  \caption{FEM}
  \label{fig:ex4-isodurfaceFEM}
\end{subfigure}
\caption{Comparison of Iso-surfaces of vertical displacement $u_{\text{z}}$ in a cube with penny-shaped crack problem. }
\label{fig:ex4isodurface}
\end{figure}

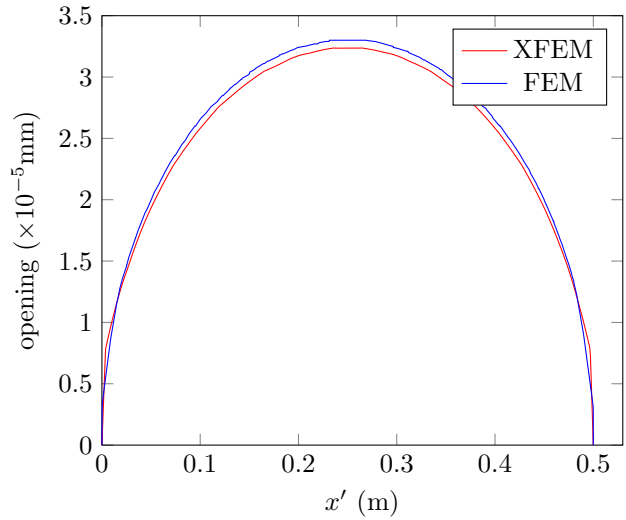
\begin{figure}
\centering
\begin{tikzpicture}
\begin{axis}[
    xlabel={ ${x}'$ (m)},
    ylabel={opening ($\times10^{-5}$mm)},
    xmin=0, xmax=0.53,
    ymin=0.0, ymax=3.5,
    xtick={0.0,0.1,0.2,0.3,0.4,0.5},
    ytick={0.0,0.5,1.0,1.5,2.0,2.5,3.0,3.5},
    legend pos=north east,
    ymajorgrids=false,
    grid style=dash,
]

\addplot[
    color=red,
    mark=false,
]
    file[] {openingXFEM3D.dat};
   \addlegendentry{XFEM}

\addplot[
    color=blue,
    mark=false,
]  
 file[]{openingFEM3D.dat};
    \addlegendentry{FEM}
\end{axis}
\end{tikzpicture}
\caption{Opening profile along the diameter of the planar penny-shaped crack; XFEM vs FEM results.}
\label{fig:ex4-opening3D}
\end{figure}

\begin{figure}
\centering
\begin{subfigure}{.5\textwidth}
  \centering
  \includegraphics[width=.95\linewidth]{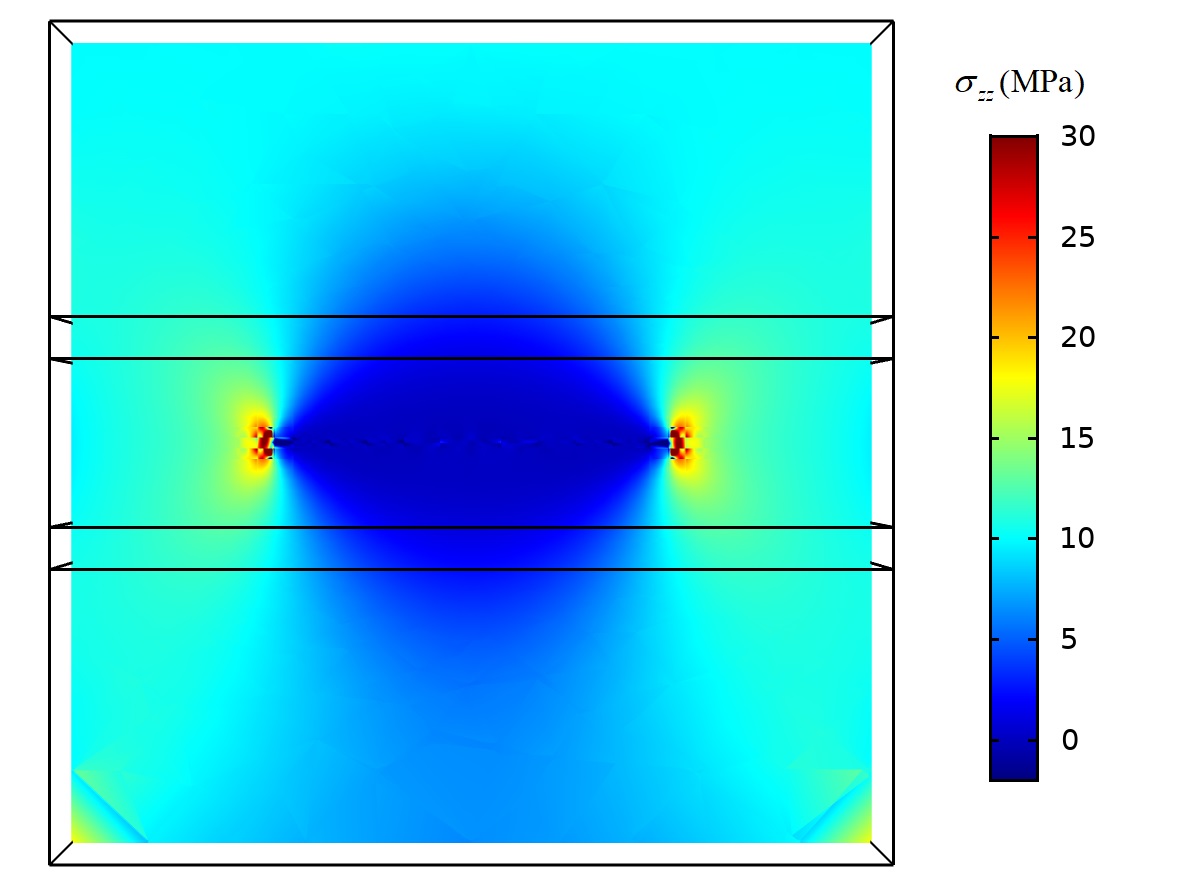}
  \caption{XFEM}
  \label{fig:ex4xfemstress}
\end{subfigure}%
\begin{subfigure}{.5\textwidth}
  \centering
  \includegraphics[width=0.95\linewidth]{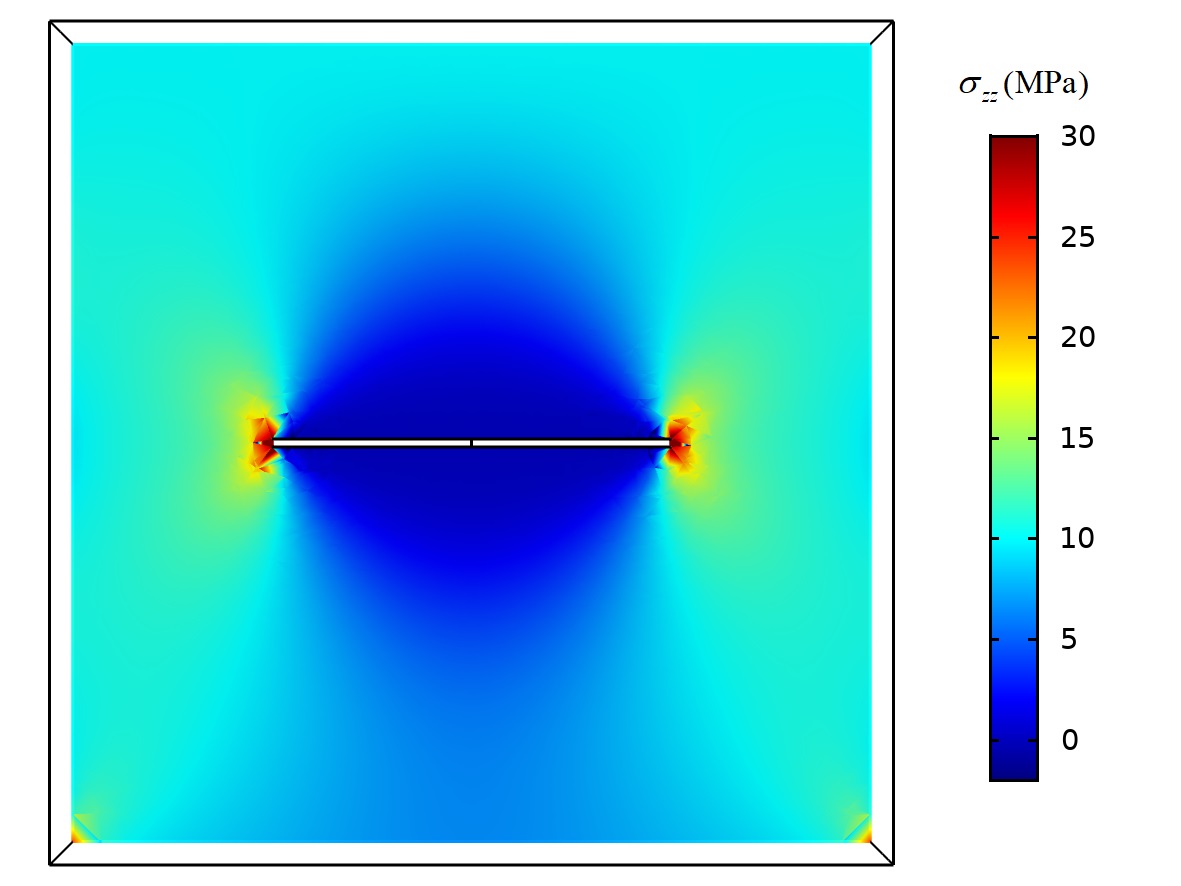}
  \caption{FEM}
  \label{fig:ex4femstress}
\end{subfigure}
\caption{Comparison of vertical stress $\sigma _{\text{zz}}$ contours in y-z plane in the cube with a penny-shaped crack problem. }
\label{fig:ex4Szz}
\end{figure}

Fig. \ref{fig:ex4isodurface} illustrates iso-surfaces of the vertical displacement $u_{\text{y}}$ contour at either sides of the crack. It is observed that the XFEM implementation of the strong discontinuity is in perfect agreement with the FEM results. The crack opening displacements associated with both models, along their diameter, are presented in Fig. \ref{fig:ex4-opening3D}. Note that the accuracy of the proposed XFEM results lies within 3\% of the FE analysis. Again, this slight discrepancy primarily pertains to the relatively negligible void thickness in the finite element simulation. Contour of the stress distributions in z-direction $\sigma_\text{zz}$ for both methods is presented in Fig. \ref{fig:ex4Szz}, which further validates the accuracy of the proposed implementation procedure.

\begin{figure}[!t]
\centering\includegraphics[width=0.5\linewidth]{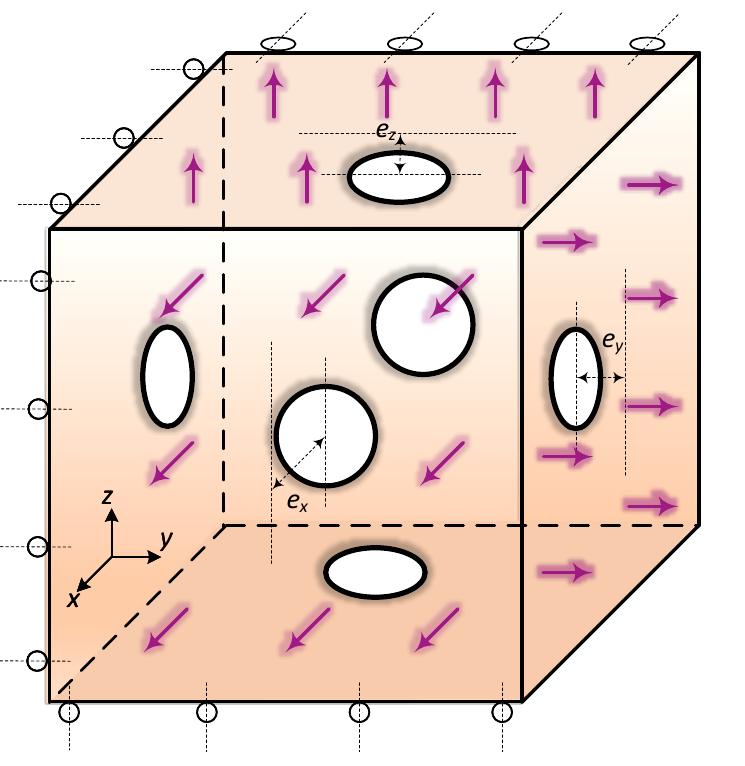}
\caption{A cube with multiple penny-shaped cracks; geometry and boundary conditions.}
\label{fig:ex-4 geometrymultiple}
\end{figure}

\begin{figure}
\centering
\begin{subfigure}{.5\textwidth}
  \centering
  \includegraphics[width=.99\linewidth]{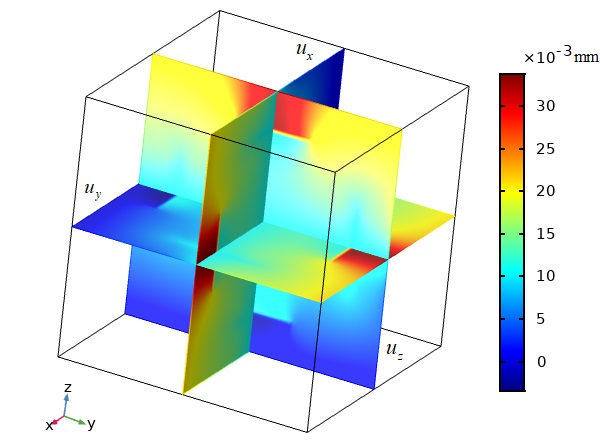}
  \caption{XFEM}
  \label{fig:ex4-uXFEMmultiple}
\end{subfigure}%
\begin{subfigure}{.5\textwidth}
  \centering
  \includegraphics[width=0.99\linewidth]{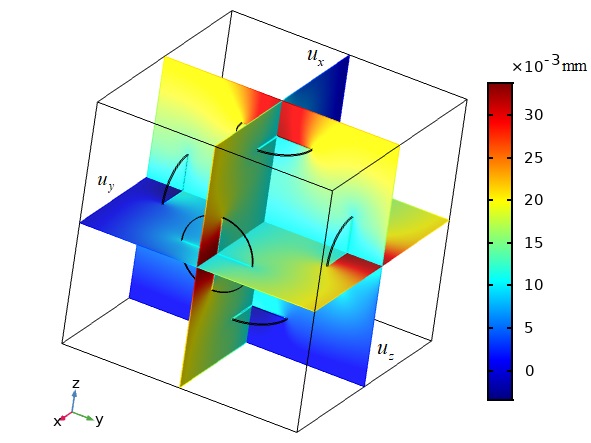}
  \caption{FEM}
  \label{fig:ex4-uFEMmultiple}
\end{subfigure}
\caption{Comparison of the displacement distribution contours for a cube with multiple penny-shaped cracks. }
\label{fig:ex4-umultiple}
\end{figure}

\begin{figure}
\centering
\begin{subfigure}{.5\textwidth}
  \centering
  \includegraphics[width=.95\linewidth]{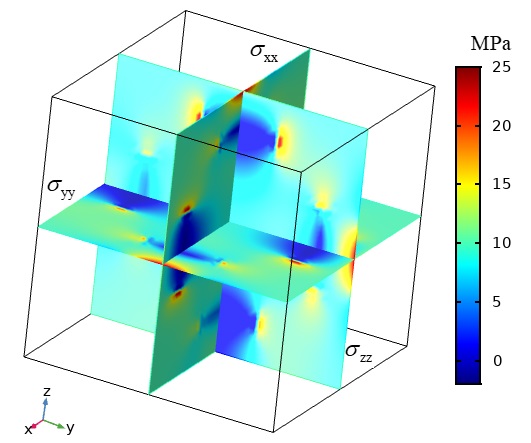}
  \caption{XFEM}
  \label{fig:ex4-SXFEMmultiple}
\end{subfigure}%
\begin{subfigure}{.5\textwidth}
  \centering
  \includegraphics[width=0.95\linewidth]{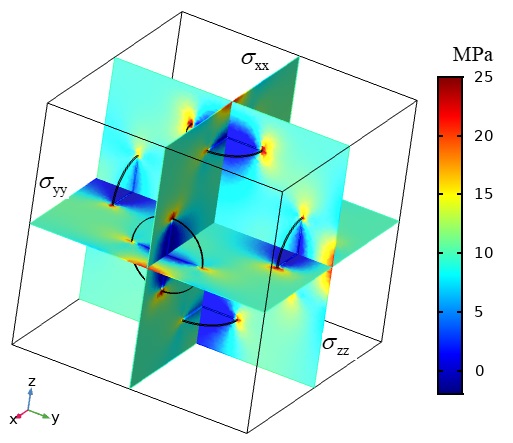}
  \caption{FEM}
  \label{fig:ex4-SFEMmultiple}
\end{subfigure}
\caption{Comparison of the distribution contours of normal components of the stress field for a cube with multiple penny-shaped cracks. }
\label{fig:ex4-Smultiple}
\end{figure}

At the end of this example, the above-mentioned three-dimensional crack tool is elaborated to simulate multiple penny-shaped cracks as depicted in Fig. \ref{fig:ex-4 geometrymultiple}. Same problem definition as the previous case is adopted except for the presence of six penny-shaped cracks of radius $0.15$ m inside the domain. The cracks are located parallel to the cube faces with $e_{\text{x}}=e_{\text{y}}=e_{\text{z}}=0.15$ m lateral distance from the boundary surfaces. Three faces of the cube are stipulated as fixed in normal direction, and the remaining three are subjected to tensile tractions with the magnitude of $10$ MPa. Contours of the displacement field as well as the orthogonal components of the stress field are respectively depicted in Figs. \ref{fig:ex4-umultiple} and \ref{fig:ex4-Smultiple}. The results confirm the flexibility of the proposed framework in handling more complex scenarios in 3D crack analysis problems.

%% file: conclusions.tex
\section{Conclusions}
\label{S:5 (Conclusions)}

In this study, an XFEM implementation in COMSOL Multiphysics is presented and applied to crack analysis in 2D and 3D solid domains. By employing a special weak form of the governing equations, the enrichment strategy is implemented within the framework of COMSOL Multiphysics software. Distinct Solid Mechanics modules are adopted to incorporate the standard and enriched parts of the displacement field in the context of XFEM. The stress intensity factor calculations, pre-processing of the model, level set updating and the crack propagation analysis are performed by means of the built-in features of the software in conjunction with external MATLAB functions. The implementational aspects and available remedies for modeling issues are explained in detail. In the first example, the accuracy of the SIF analysis of the proposed implementation is validated against benchmark analytical solutions in 2D settings. The second example is devoted to highlight the capability of the proposed strategy in dealing with heavily fractured bulks. Next, two crack growth studies, involving single and multiple crack propagation, are presented to demonstrate the capabilities of the extended framework in cases where the geometry is subject to changes. In the final example, an extension of the proposed procedure for modeling single/multiple cracks in 3D domains is carried out. In all numerical examples, the results obtained indicate excellent agreement with the existing analytical/computational solutions or experimental measurements. This implies the soundness of the proposed implementation strategy and its tremendous potential for modeling fractures. Future developments could be aimed at complex multi-field problems, which is the cornerstone of COMSOL Multiphysics package.